\documentclass[prl,twocolumn,aps]{revtex4}

\usepackage{amsmath,amssymb}
\usepackage{latexsym}
\usepackage{graphicx}

\def\rootfig{./fig-ps/}

\begin{document}
\title{Dynamics of Few Co-rotating Vortices in Bose-Einstein Condensates}

\author{
R. Navarro$^1$,
R.\ Carretero-Gonz\'{a}lez$^1$,
P.J.\ Torres$^2$,
P.G.\ Kevrekidis$^3$,
D.J.\ Frantzeskakis$^4$,
M.W.\ Ray$^5$,
E. Altunta\c{s}$^{5}$, and
D.S.\ Hall$^5$
}

\affiliation{%
$^1$Nonlinear Dynamical Systems Group,
Computational Science Research Center, and
Department of Mathematics and Statistics,
San Diego State University, San Diego,
CA 92182-7720, USA
\\
$^2$Departamento de Matem\'atica Aplicada,
Universidad de Granada, 18071 Granada, Spain
\\
$^3$ Department of Mathematics and Statistics, University of
Massachusetts, Amherst MA 01003-4515, USA
\\
$^4$ Department of Physics, University of Athens, Panepistimiopolis,
Zografos, Athens 157 84, Greece
\\
$^5$ Department of Physics, Amherst College,
Amherst, Massachusetts, 01002-5000 USA
}

\begin{abstract}
\label{ssec:abs}
We
study the dynamics of small vortex clusters with few (2--4)
co-rotating vortices in Bose-Einstein condensates
by means of experiments, numerical computations, and theoretical analysis.
All of these approaches corroborate
the counter-intuitive presence of a dynamical instability of
symmetric vortex configurations.
The instability arises as a pitchfork bifurcation at sufficiently large
values of the
angular momentum that induces the emergence and
stabilization of {\it asymmetric} rotating vortex configurations.
The latter are quantified in the theoretical model and
observed in the experiments. The dynamics is explored
both for the integrable two-vortex system, where a reduction
of the phase space of the system provides valuable insight, as well as for
the non-integrable three- (or more) vortex
case, which additionally
admits the possibility of chaotic trajectories.
\end{abstract}

\maketitle

{\it Introduction.}  The realm of atomic Bose-Einstein condensates (BECs)
\cite{bec} has offered a pristine setting for
studies on the dynamics of few-vortex clusters~\cite{castin}.
Most investigations, however, have  focused on
either a single vortex or
large scale vortex
lattices~\cite{Fetter2001,Kevrekidis2008,Fetter2009,Newton2009}.
Recently, theoretical investigations
on the study of clusters of 2--4
vortices~\cite{Crasovan2002,Crasovan2003,Zhou2004,Mottonen2005,Pietila2006,Li2008,Middelkamp2010,Kuopanportti2011,Torres2011},
have appeared, chiefly
motivated by the experimental realizations
of such states~\cite{Neely2010,Freilich2010,Seman2010,Middelkamp2011}.
This focus has been heretofore centered on the fundamental building
block of the vortex dipole, i.e., a pair of counter-circulating vortices.

Our aim in the present work, in contrast, 
is to explore the dynamics of
small vortex clusters of 2--4 vortices
that belong to the co-rotating (same charge) variety.
The original work of Ref.~\cite{castin} and subsequent efforts~\cite{Afta03_04_Danaila05}
have already paved the way for an understanding of symmetric few-vortex
configurations rotating 
as a rigid body, 
and their three-dimensional
generalizations, i.e., 
U- and S-shaped vortices, as well as
vortex rings~\cite{komineas_rev}. In this context, our work
presents a rather unexpected twist: we have found that,
under suitable conditions, the
usual symmetric,
co-rotating
vortex configurations (centered line, triangle, and square)
become {\it dynamically unstable}. 
More specifically,
these states become subject to symmetry breaking, pitchfork
bifurcations that lead to the spontaneous emergence of {\it stable
asymmetric rotating vortex clusters}.

We present our analysis of these features in the integrable
(at the reduced particle level)
setting of a co-rotating vortex pair, and illustrate their generality by
further considering a rigidly rotating vortex triplet and quadruplet. In the first case, we devise
a theoretical formulation that not only explores the instability
and manifests its growth rate, but also enables a visualization
of a two-dimensional reduced phase space of the system in which
the  pitchfork bifurcation
becomes transparent.
In the latter cases, we 
suitably parametrize the system, exploring
the different regimes of symmetric and asymmetric periodic orbits.
Our theoretical analysis treats vortices as classical particles,
with dynamics
governed
by ordinary differential equations (ODEs).
This reduction of the original
vortex cluster system
allows for the analytical characterization, numerical observation and
experimental
confirmation of the symmetry breaking phenomena.

{\it Theoretical Analysis.}
As illustrated in Refs.~\cite{Middelkamp2011,Torres2011}, and justified
by means of a variational approximation~\cite{pelin2011}, vortex dynamics
governed by the two-dimensional mean-field Gross-Pitaevskii equation,
%
\begin{eqnarray}
i \partial_t \psi =-\frac{1}{2} \Delta \psi + \frac{1}{2}\Omega^2(x^2 + y^2) \psi + |\psi|^2 \psi,
\label{pde1}
\end{eqnarray}
can be reduced to a system of ODEs for the vortex positions. In the original partial
differential equation (PDE) model (\ref{pde1}), the
time is measured in units of $\omega_z^{-1}$, while the positions
are measured in units of harmonic oscillator length along the
$z$-direction and $\Omega=\omega_x/\omega_z=\omega_y/\omega_z$,
with $\omega_j$ being the harmonic trap frequency
along the $j$-direction (see, e.g., Ref.~\cite{Kevrekidis2008}).
This
ODE reduction is 
the starting point for our analysis in the co-rotating case.

The dynamics of vortex $m$ at position $(x_{m},y_{m})$
arises from two contributions: (i) a position-dependent
vortex precession about the trap center
with frequency $S_m\,\omega_{\rm pr}$, and (ii) a vortex-vortex interaction
with vortex $n$ that induces a velocity perpendicular to their line of sight of
magnitude $S_n\,\omega_{\rm vort}/\rho_{mn}^2$, where $\rho_{mn}$ is
the distance between vortices $m$ and $n$, $S_m$ and $S_n$ are
their respective charges, and $\omega_{\rm vort}$ 
is a dimensionless constant; see Ref.~\cite{Middelkamp2011,Torres2011}.
The equations governing the dynamics of $N$ interacting vortices embedded in
a condensate are therefore
\begin{equation}
\label{middlecamp}
\begin{array}{rcl}
\dot{x}_{m} &=&
\displaystyle
-S_{m}\omega _{\mathrm{pr}}y_{m}-\frac{\omega_{\rm vort}}{2}\sum_{n\neq m}S_{n}%
\frac{y_{m}-y_{n}}{\rho _{mn}^{2}},  \\[3.5ex]
\dot{y}_{m} &=&
\displaystyle
\phantom{-}S_{m}\omega_{\mathrm{pr}}x_{m}+\frac{\omega_{\rm vort}}{2}\sum_{n\neq m}S_{n}%
\frac{x_{m}-x_{n}}{\rho _{mn}^{2}}.
\end{array}
\end{equation}
The precession about the trap center can be approximated by
$\omega_{\rm pr}={\omega_{\rm pr}^{0}}/({1-{r^{2}}/{R_{\rm TF}^{2}}})$, where
the frequency at the trap center is
$\omega _{\rm pr}^{0}=\ln \left( A\frac{\mu }{\Omega }\right)/R_{\rm TF}^{2}$,
$\mu$ is the chemical potential, $R_{\rm TF}=\sqrt{2 \mu}/\Omega$ is the
Thomas-Fermi (TF) radius, and $A$ is
a numerical constant~\cite{Fetter2001,Middelkamp2011,Torres2011}.
To describe better the actual vortex dynamics in the trap,
the constant $\omega_{\rm vort}$ in Eqs.~(\ref{middlecamp})
may be adjusted to
account for the screening of vortex interactions due to the
background density modulation~\cite{screening}.

\begin{figure*}[tbph]
\begin{center}
\includegraphics[width=3.5cm,height=3.1cm]{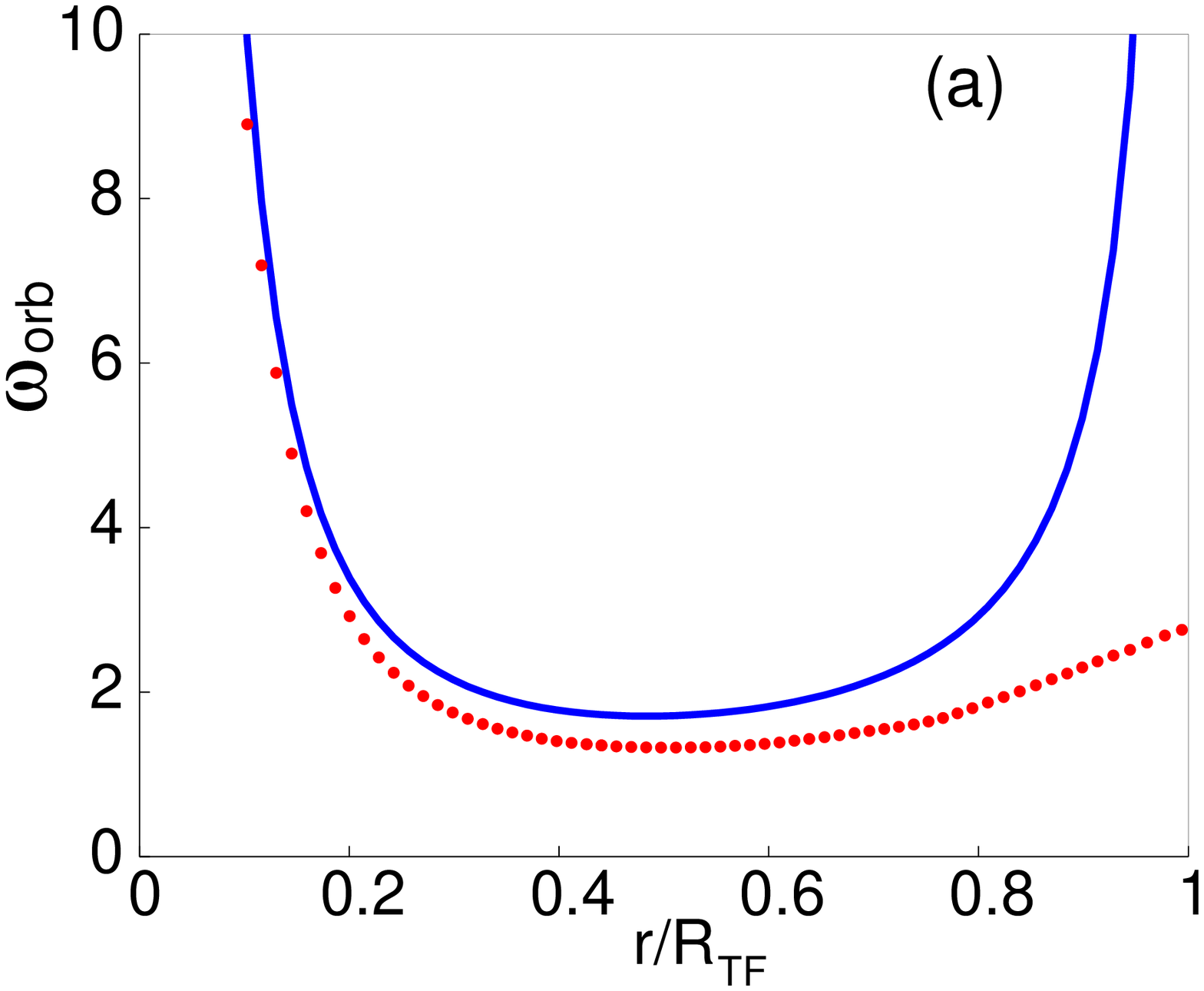}
\includegraphics[width=3.5cm,height=3.1cm]{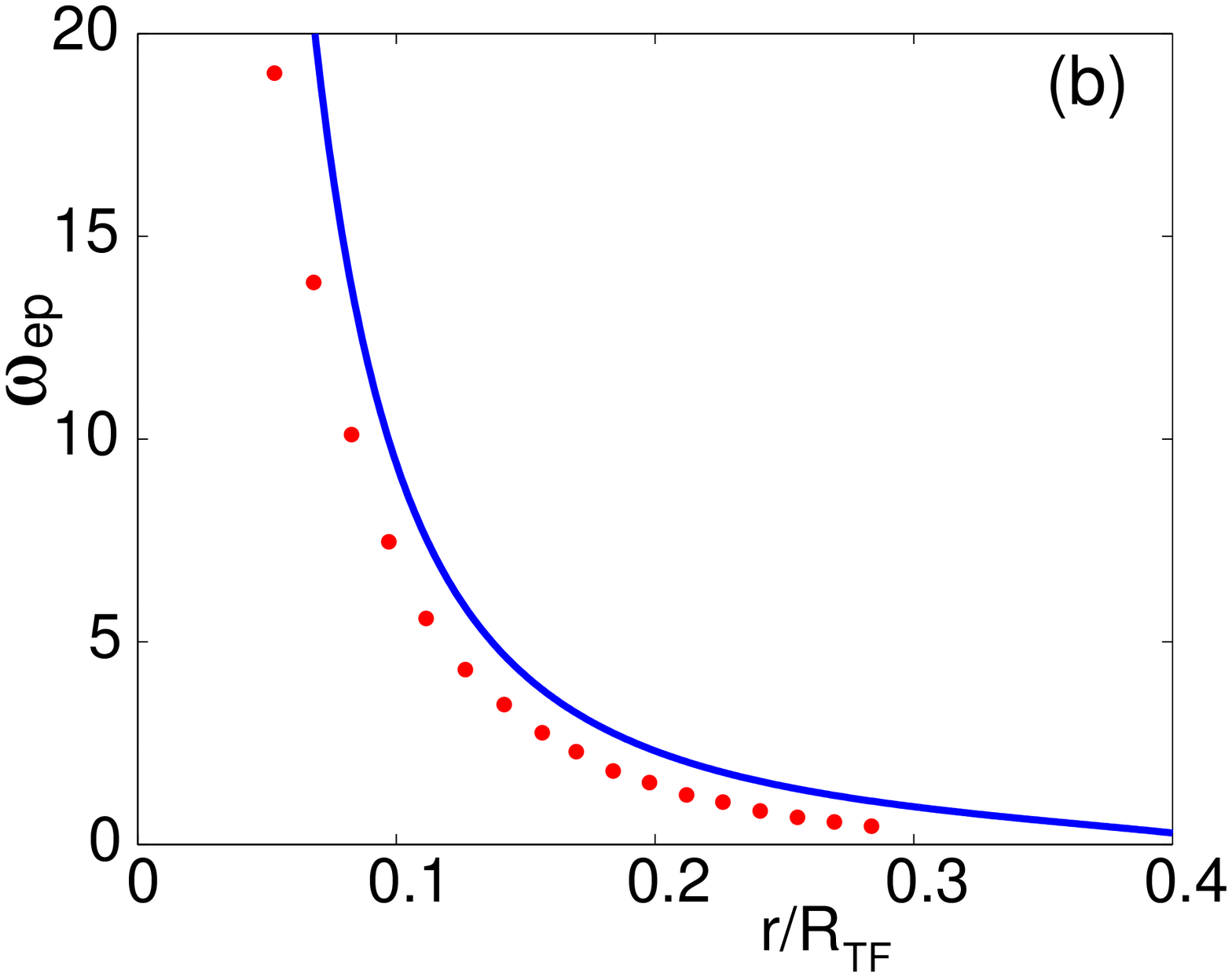}
\includegraphics[height=3.0cm]{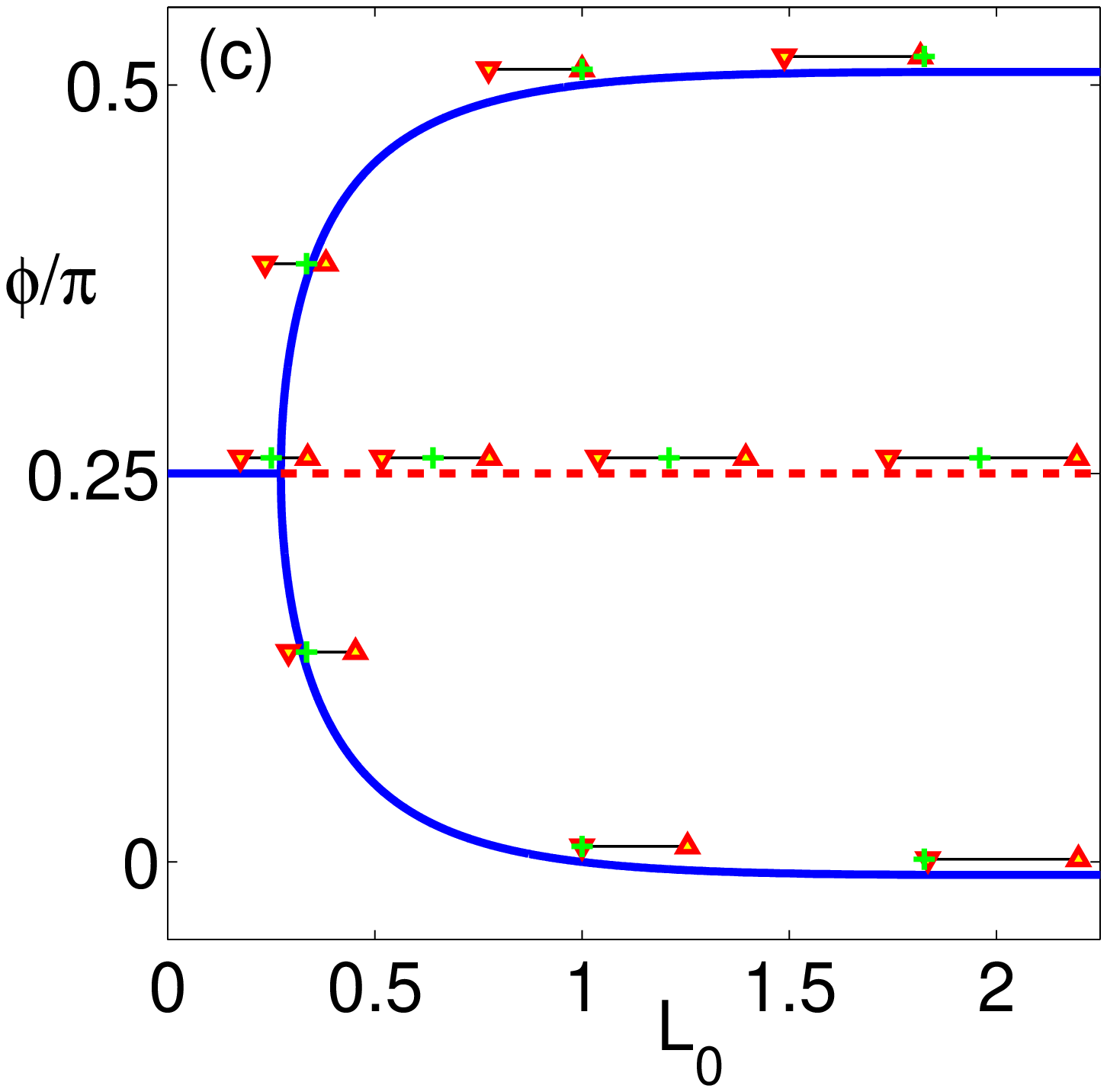}
\includegraphics[height=3.0cm]{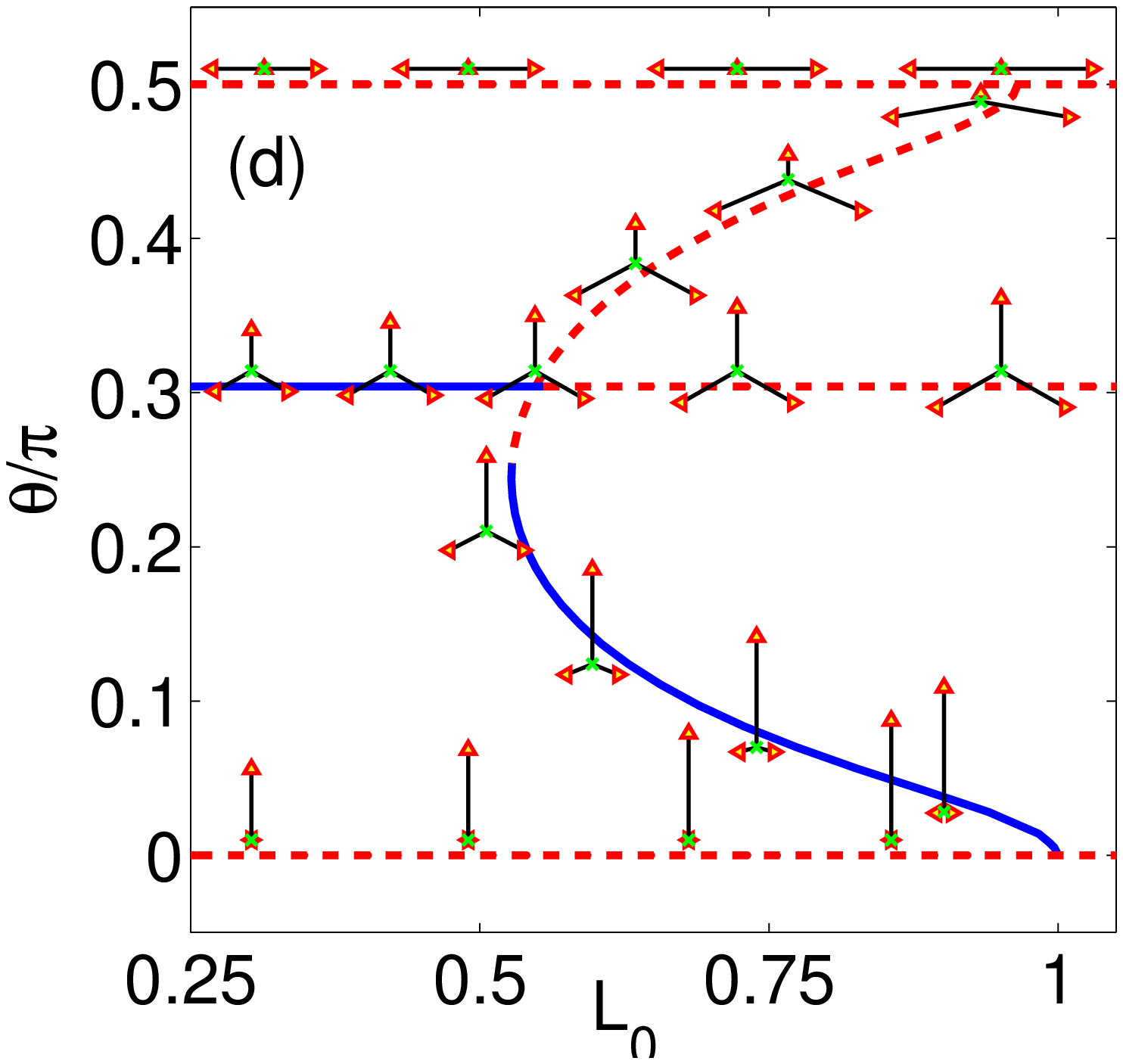}
\includegraphics[height=3.0cm]{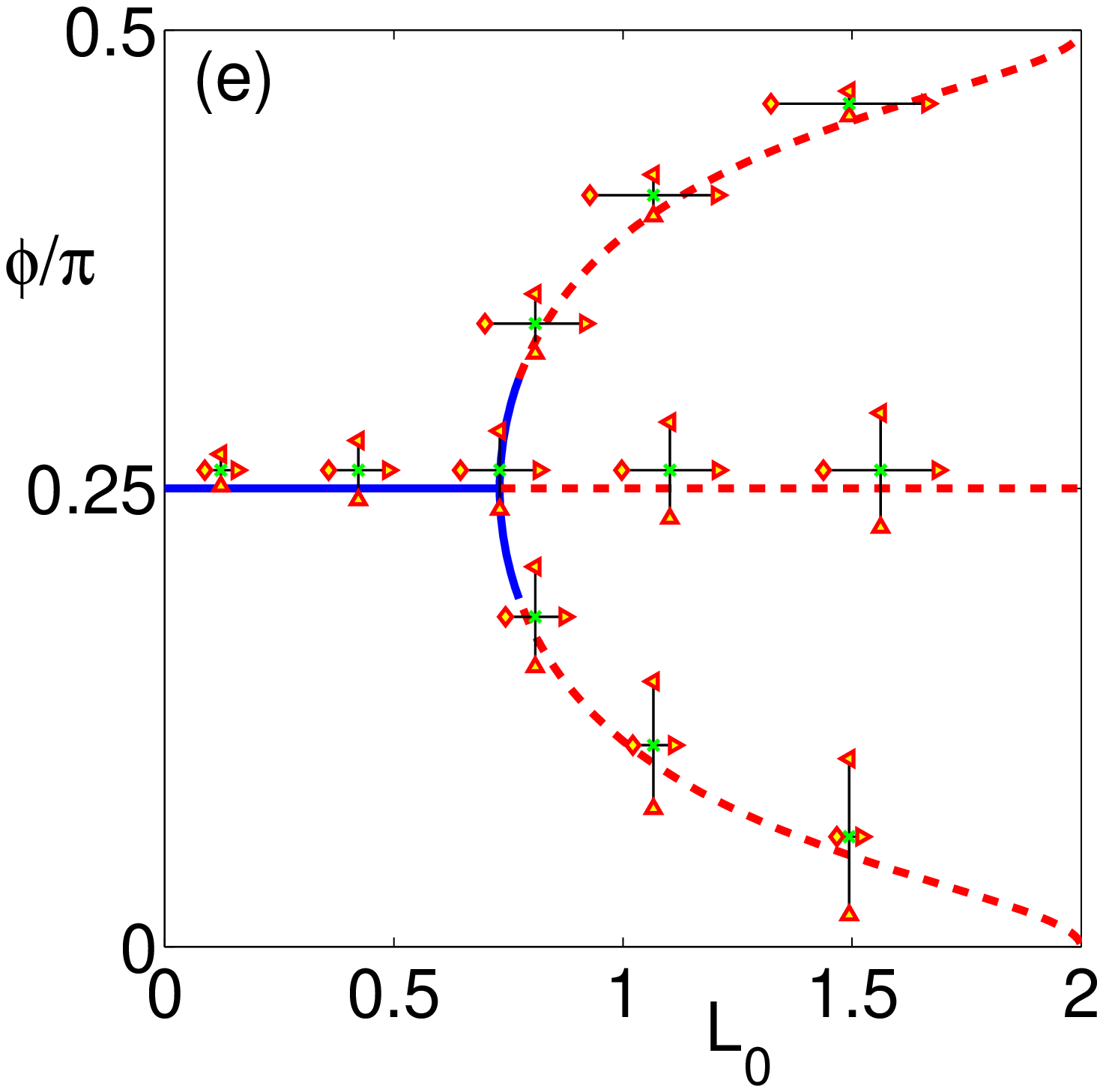}
\end{center}
\vspace{-0.4cm}
\caption{(Color online)
(a) Orbital and (b) epitrochoidal frequency as a function of the
radial position from the trap's center for two vortices. Both frequency and
radial position are in rescaled units.
The solid line represents results from the
ODE and the dotted line from the PDE.
The vanishing of the latter signals the onset of instability.
Here, $\Omega =0.05$ and $\protect\mu %
=1$.
Panel (c) depicts the quantity $\phi/\pi$,
which equals 1/4 when $r_1=r_2$, as a function
of the square root of the angular momentum for $c=0.1$.
Panels (d) and (e) depict the corresponding phenomena for
$N=3$ and $N=4$ vortices for $c=0.1$.
Panels (c), (d) and (e) include a few configurations along
the main bifurcation branches ([blue] solid and [red] dashed lines
corresponding, respectively, to stable and unstable configurations) 
depicting the relative position
of the vortices (red triangles) with respect to the center of the condensate
(green crosses).
%
}
\label{pde_per}
\end{figure*}

\begin{figure}[tbph]
\begin{center}
\includegraphics[width=4cm,height=3cm]{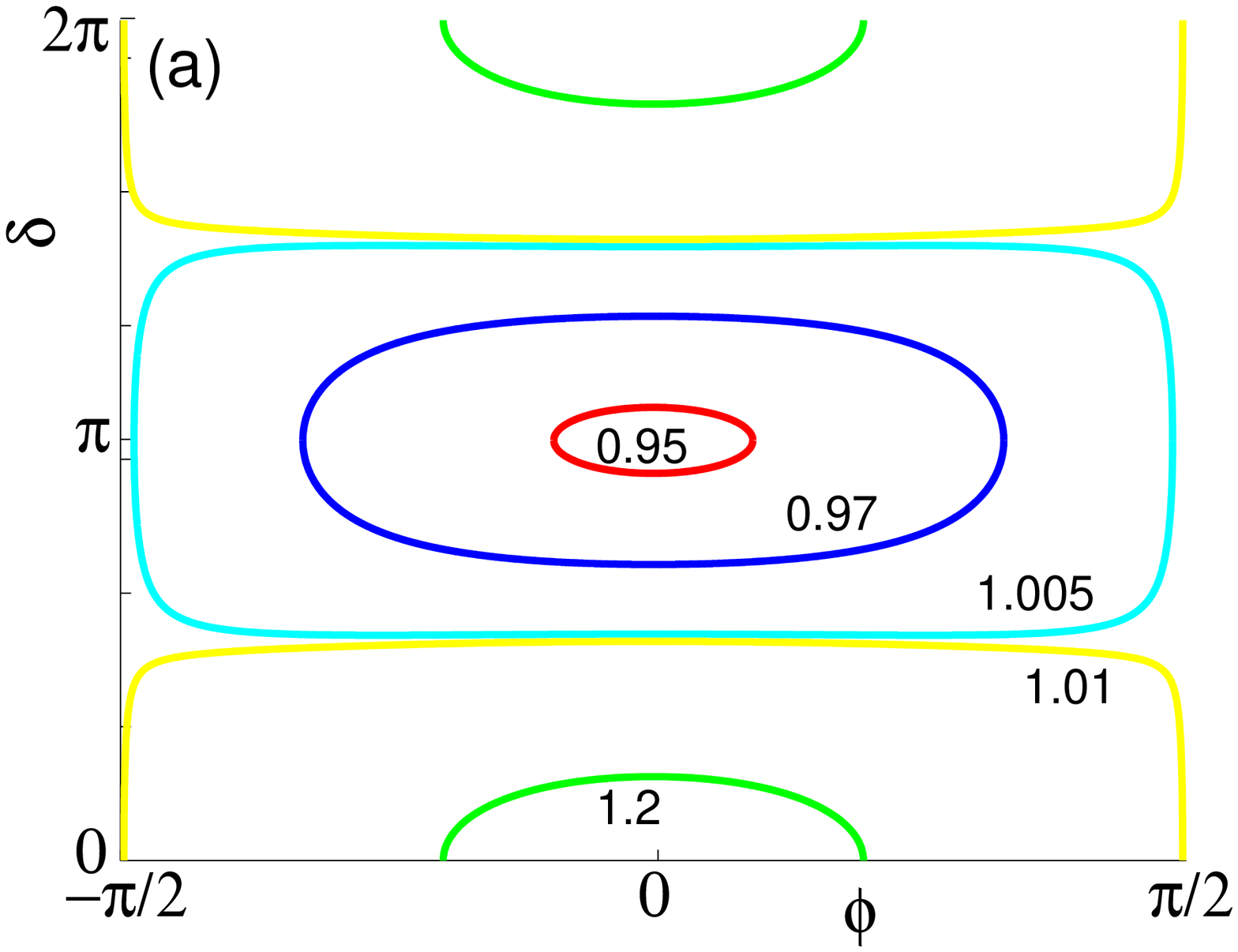}
~
\includegraphics[width=4cm,height=3cm]{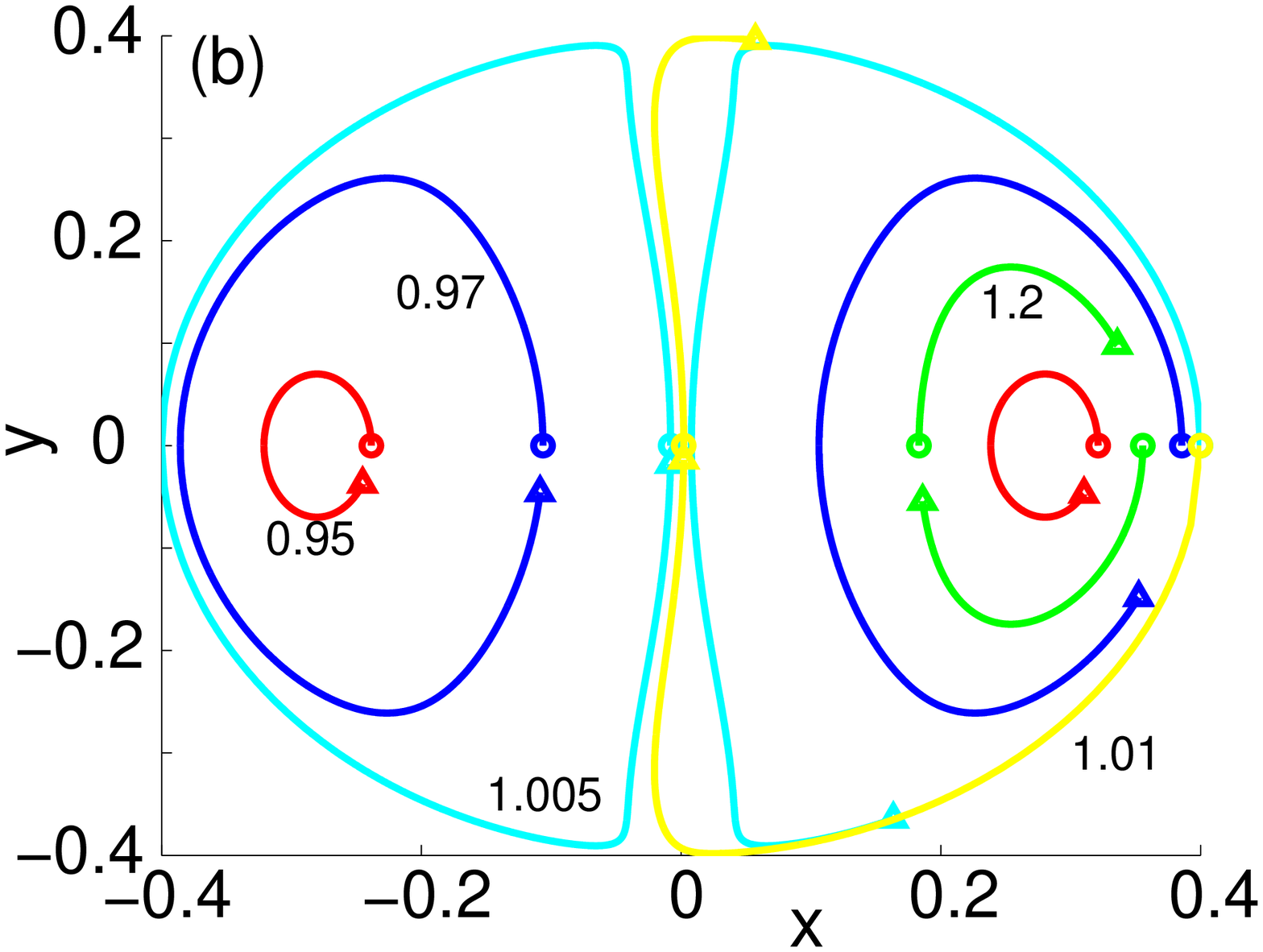}\\[0.5ex]
\includegraphics[width=4cm,height=3cm]{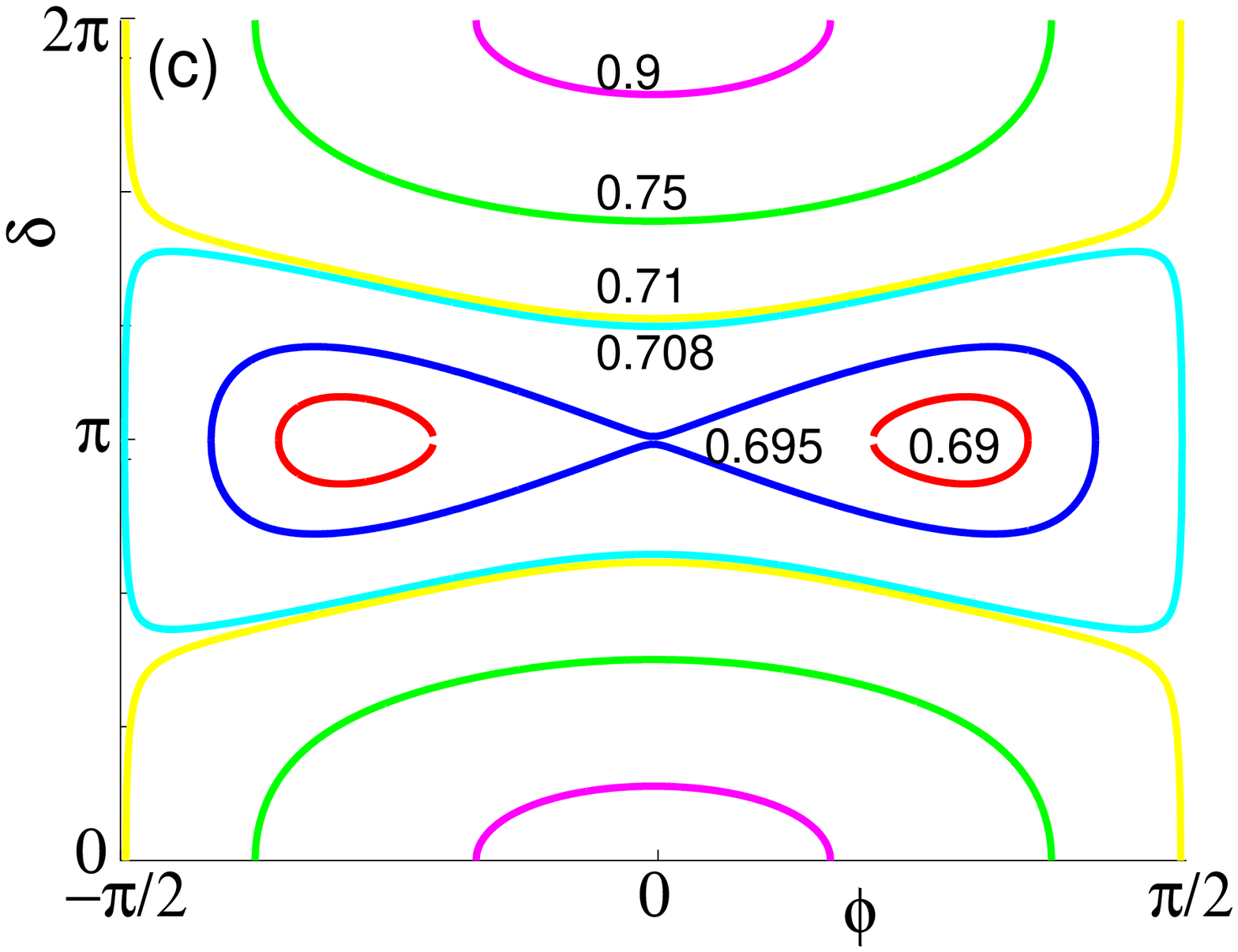}
~
\includegraphics[width=4cm,height=3cm]{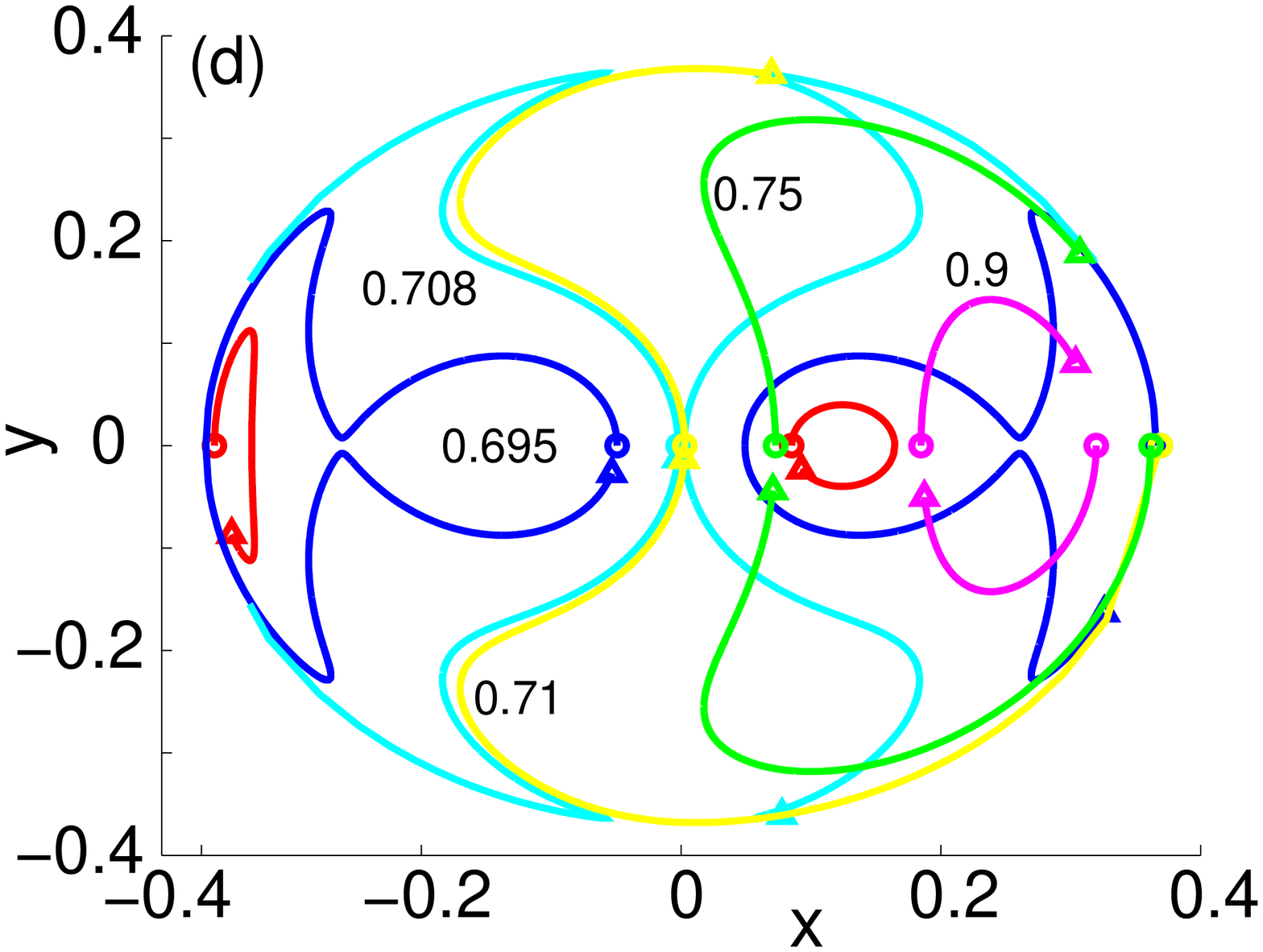}\\
\end{center}
\vspace{-0.5cm}
\caption{(Color online)
(a) Contours from the reduced Hamiltonian for vortices close to the
center of the trap, i.e.~$L_0=0.16$
below critical ($L_0<L_{\rm cr}=0.273$),
and (b) their corresponding orbits.
Panels (c) and (d) depict, respectively, the same quantities but
for vortices further out from the center for $L_0=0.36$
above critical ($L_0>L_{\rm cr}=0.273$).
The circles and triangles correspond, respectively, to the initial and final
positions of the two vortices. Each color on the contour plot matches the
corresponding orbit in the position diagram. Here, $c=0.1$.}
\label{hamR02}
\end{figure}

We now focus on the case of two identical vortices of unit charge
$S_1=S_2=1$. We proceed to adi\-men\-sio\-na\-li\-ze Eqs.~(\ref{middlecamp})
by scaling $(x,y)$ by $R_{\rm TF}$ and
time by $1/\omega_{\rm pr}^{0}$, and use
polar coordinates $(x_n,y_n)=(r_n \cos(\theta_n), r_n \sin(\theta_n))$.
%
%
We then seek symmetric stationary states $r_{1}=r_{2}=r_{\ast}$
and $\theta _{1}-\theta _{2}=\pi$, and find the following
frequency of the co-rotating vortices:
\begin{equation}
\omega_{\rm orb}= \dot{\theta}_{1}=\dot{\theta}_{2}=\frac{c}{2r_{\ast }^{2}}+%
\frac{1}{1-r_{\ast }^{2}},
\label{worbeq}
\end{equation}
where $c=\frac{1}{2}({\omega _{\mathrm{vort}}}/{\omega _{\mathrm{pr}}^{0}})$
yields a measure of the relative strength of vortex interaction and
spatial inhomogeneity.
The comparison of the orbital frequency between the ODE and the PDE
models  is given in Fig.~\ref{pde_per}a.
Given the co-rotating nature of this state, consideration of
$\delta_{mn}=\theta_m-\theta_n$ renders this state a stationary
one; linearizing around it using
$r_{m}=r_{\ast }+R_{m}$ and $\delta _{mn}=\pi +\delta _{m}$ yields the
following equations of motion for the perturbations about the
symmetric
equilibrium:
%
%
\begin{equation*}
\ddot{R}_{m} =-\frac{\omega _{{\rm ep}}^{2}}{2}\left(
R_{n}-R_{m}\right),
\quad
\ddot{\delta}_{m} =-\frac{\omega _{{\rm ep}}^{2}}{2}\left( \delta
_{m}-\delta _{n}\right),
\end{equation*}
with
$\omega _{{\rm ep}}^{2}=\frac{c^{2}}{2r_{\ast }^{4}}-\frac{2c}{^{\left(
1-r_{\ast }^{2}\right) ^{2}}}$. It is then straightforward to
observe that this squared epitrochoidal (motion of a point
in a circle that is rotating about another circle)
relative precession frequency of the two vortex positions and phases
{\it changes sign} at
$r_{\rm cr}^2 = \sqrt{c}/(\sqrt{c}+2)$. This signals our first
fundamental result, namely the {\it destabilization} of the
symmetric 2-co-rotating vortex state for sufficiently large
symmetric distances of the vortices from the trap center.
A comparison of the ODE and PDE models for the orbital and epitrochoidal
precession frequencies for these two cases is given in Fig.~\ref{pde_per}a-b.

The dynamical instability of symmetric states
suggests the potential existence of additional, asymmetric, ones.
Seeking states with $\delta_{mn}=\pi$ and $r_1^{\ast} \neq r_2^{\ast}$ yields
\begin{equation*}
-r_{1}^{\ast }r_{2}^{\ast }(r_{1}^{\ast }+r_{2}^{\ast })^{2}+c\left(
1-r_{1}^{\ast 2}\right) \left( 1-r_{2}^{\ast 2}\right) =0,
\end{equation*}
which will be the condition defining our radially {\em asymmetric} co-rotating
solutions. The mirror symmetry of the 2-vortex system predisposes
towards the pitchfork, symmetry breaking nature of the relevant
bifurcation, a feature verified by the diagram
of Fig.~\ref{pde_per}c. This diagram is given for the angle
$\phi=\tan^{-1} r_{2}/r_{1}$
as a function of the angular momentum
$L_{0}=r_{1}^{2}+r_{2}^{2}$, which is a conserved quantity for
our system.
It is interesting to note that if the single dimensionless parameter of
the system is small ($c<3$), then the critical value $L_{\rm cr}$ for
$L_0$ ---at which
the bifurcation from symmetric to asymmetric periodic orbits occurs---
is supercritical, while if $c$ is sufficiently large ($c>3$), it becomes
subcritical \footnote{It should be noted here, however, that the latter
bifurcation is merely of mathematical interest, as for typical values
of the physical parameters, the supercritical scenario is the one
which is realized.} (not shown).

To elucidate the pitchfork nature of the bifurcation, we develop a
phase plane representation for all 2-vortex configurations.
The integrability of the reduced 2-particle description
can be understood on the basis
of the fact that this 4-dimensional system has two integrals of
motion,
namely the angular momentum $L_0$, 
defined above, and the Hamiltonian $H$,
which can be written in polar coordinates as
%
\begin{equation*}
H=\frac{1}{2}\ln \left[ \left( 1-r_{1}^{2}\right) \left( 1-r_{2}^{2}\right)
\right]-\frac{c}{2}\ln \left[ r_{1}^{2}+r_{2}^{2}-D) \right],
\end{equation*}
where $D\equiv 2r_{1}r_{2}\cos(\delta)$ and $\delta=\theta_2-\theta_1$.
Using $L_0$ and the angle $\phi$ 
to express $r_1$ and $r_2$, one can
rewrite the Hamiltonian as a function of $(\phi,\delta)$
%
having thus effectively reduced the 4-dimensional system into a 2-dimensional
one.
Thus, for different values of $L_0$, we can represent the orbits in the
effective phase
plane of $(\phi,\delta)$ in which the different orbits correspond to
iso-energetic $H(\phi,\delta)$=const.\ contours.
This is done in Fig.~\ref{hamR02} for values that are both
below and above than the critical value of $L_0$ at fixed $c$.
It
can then be inferred that the symmetric fixed point with $(\phi,\delta)=(\pi/4,\pi)$
is stable in the former case, while it destabilizes in the latter case
through the emergence of two additional asymmetric ($\phi \neq
\pi/4$) states along the horizontal line $\delta=\pi$ of
anti-diametric vortex states.

Remarkably,  although the properties of the system dramatically change
as we go from two vortices to three and four, the
symmetry breaking bifurcation associated with the symmetric solutions persists.
In particular, when $N>2$, the persistence of the two conservation
laws discussed above is not sufficient to ensure integrability of the
system, and its absence is manifested in a dramatic form in the resulting
6 ($N=3$) and 8 ($N=4$) dimensional
systems through the presence of chaotic orbits.
Nevertheless, one can still theoretically analyze the highly symmetric
co-rotating states of the system.

For $N=3$, this
state is an equilateral
triangle such that $r_{1}=r_{2}=r_{3}=r_{\ast }$ and $\delta _{i,i+1}=2\pi /3$,
with an orbital frequency predicted as
$\omega_{\rm orb, 3}=\frac{c}{r_{\ast }^{2}}+\frac{1}{1-r_{\ast
}^{2}}$. In the co-rotating frame, the linear stability analysis
around this rigidly rotating triangle can be performed giving
rise to an epitrochoidal frequency
$
%
\omega_{\rm ep,3}^{2}=\frac{c^{2}}{r_{\ast }^{4}}-\frac{2c}{\left( 1-r_{\ast
}^{2}\right) ^{2}}.
%
$
In this case too, a critical radius exists $r_{\rm cr,3}^2 =
\sqrt{c}/(\sqrt{c}+\sqrt{2})$, such that the symmetric state is
destabilized and asymmetric orbits arise and
are stable past this critical point as can be seen in Fig.~\ref{pde_per}d.
The dynamical picture is considerably more complicated
but the conservation of the angular momentum ensures that the dynamical
evolution resides on the surface of a Bloch sphere.
We thus define two angular variables $\tan \phi =r_{2}/r_{1}$ and
$\cos \theta =r_{3}/\sqrt{L_{0}}$ and depict the associated pitchfork bifurcation
in Fig.~\ref{pde_per}d for the subspace of solutions constrained to
$r_1=r_2$ and $\delta_{12}=\delta_{23}$.
This bifurcation diagram describes a vortex configuration containing
a stable symmetric rotating triangle before the bifurcation and
stable asymmetric rotating triangles after the bifurcation.
In addition to the equilibrium and near-equilibrium
orbits, we  observe chaotic orbits arising both in a more localized
form, exploring the vicinity of equilibrium orbits, and in a more
extended one spanning all space
(not shown).

While the general phenomena
for $N=4$
are already rather complex, some basic features can still be inferred
and the symmetry breaking nature of the proposed instability
persists
---cf. Fig.~\ref{pde_per}e. Here, $\phi=\tan^{-1} r_3/r_1$,
and we have constrained the vortices to be in a
cross with right angles and $r_1=r_3$ and $r_2=r_4$.
A general expression for the orbital
frequency of the rigidly rotating state
is
$\omega _{{\rm orb},N}=\frac{\left( N-1\right) c}{2r_{\ast }^{2}}+%
\frac{1}{1-r_{\ast }^{2}}$, which is
valid for any 
$N$. In the case of the square configuration with $r_{i}=r_{\ast }$ and
$\delta _{i,i+1}=\pi /2$, there emerge
two epitrochoidal vibrational motions with frequencies
$\sqrt{-\lambda_1}$ and $\sqrt{-\lambda_2}$, where
%
$\lambda_{1}=\frac{3c}{\left( 1-r_{\ast }^{2}\right) ^{2}}
-\frac{9c^{2}}{4r_{\ast }^{4}}$,
and
$\lambda_{2}=\frac{4c}{\left( 1-r_{\ast }^{2}\right) ^{2}}
-\frac{2c^{2}}{r_{\ast }^{4}}$.
These, in turn, correspond to two critical points: one identical to the
one given above for the $N=3$ case, and one that is always higher,
given by $r_{\rm cr,4}^2 =
\sqrt{3 c}/(\sqrt{3 c}+2)$; hence, the same phenomenology persists.

{\it Experimental Observations.}
We now briefly discuss experimental manifestations
of the symmetry breaking events discussed above and of the emergence
of asymmetric configurations.
%

The details of the
experimental setup may be found elsewhere~\cite{Freilich2010,Middelkamp2011}.
We begin with a magnetically-trapped BEC of $N \sim 5$--$8\times 10^5$ atoms in the $|F=1,m_F=-1\rangle$
hyperfine level of $^{87}$Rb. The radial and axial trap frequencies are
$(\omega_r,\omega_z)/2 \pi =(35.8,101.2)$\,Hz. Vortices are introduced through
a process of elliptical magnetic trap distortion and rotation~\cite{Hodby2001}
during evaporation~\cite{Madison2000}. In terms of the trap frequencies along the major
and minor axes of the distorted potential, $\omega_x$ and $\omega_y$ respectively,
an ellipticity $\epsilon=(\omega_x^2-\omega_y^2)/(\omega_x^2+\omega_y^2) = 0.20$ and a
rotation frequency of 8.5\,Hz usually produces a co-rotating pair. Higher rotation frequencies
are used to generate larger numbers of co-circulating vortices.

\begin{figure}[t]
\begin{center}
\includegraphics[width=8.0cm]{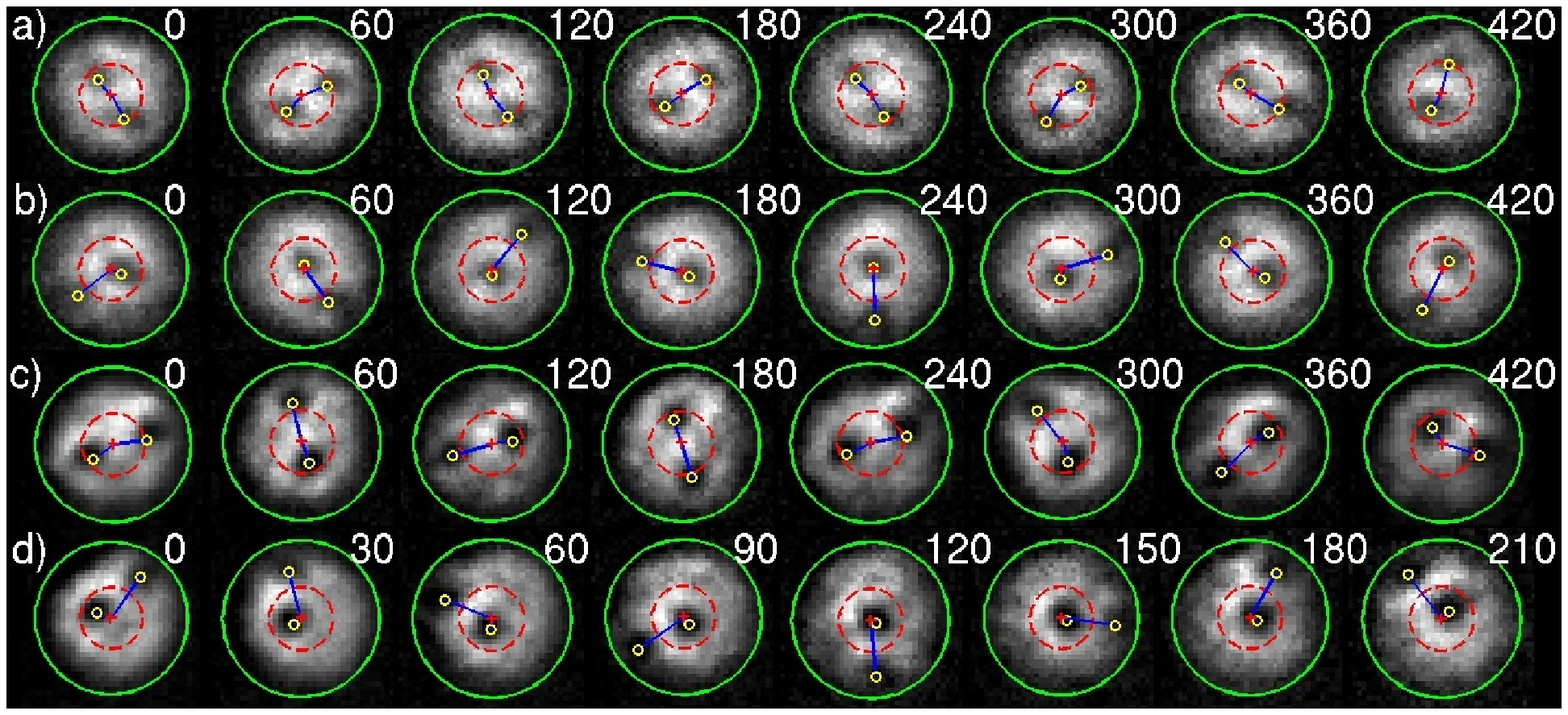}
\includegraphics[width=8.5cm,height=9.4cm]{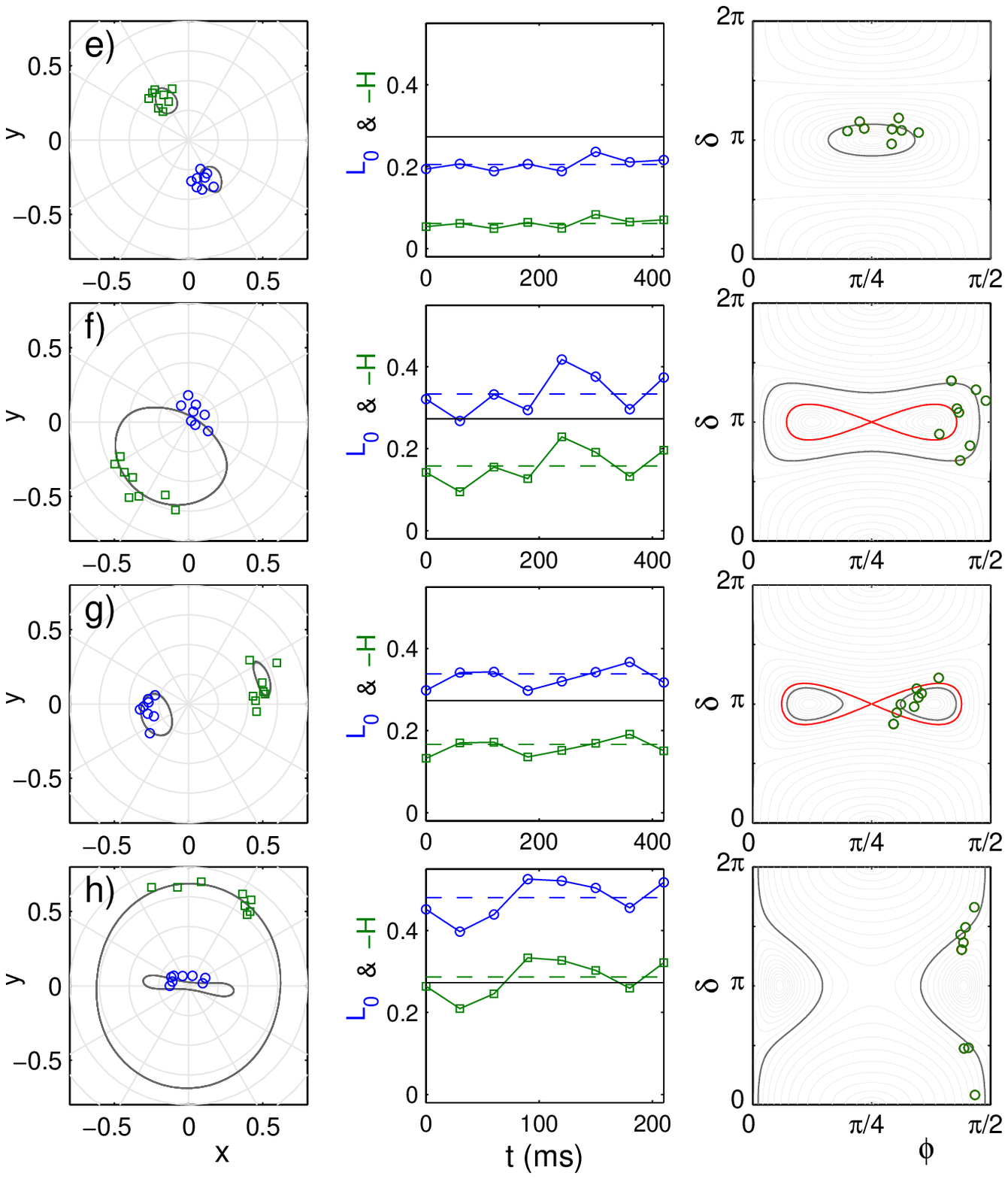}
\end{center}
\par
\vspace{-0.4cm}
\caption{(Color online)
a)--d)
Typical experimental series for the dynamics of two co-rotating vortices
(time indicated in ms).
The large (green) circles and the (red) crosses represent, respectively,
the fitted TF radius and center of the
cloud while the small (yellow) dots depict the fitted vortex centers.
The (red) dashed circles represent the critical radius, $r_{\rm cr}$,
above which symmetric orbits become unstable.
e)--h)
Manifestation of the pitchfork bifurcation for the
experimental series depicted in panels a)--d)
which correspond to $c=0.1$.
Left column: experimental vortex positions
and their corresponding orbit from the
reduced ODE model (solid line), in TF units in the
co-rotating frame.
Middle column: corresponding $L_0$ (blue circles) and $-H$
(green squares) and their averages (horizontal dashed lines) as well as
the critical value for $L_0$ (solid horizontal line).
Right column: corresponding orbits in the
$(\phi,\delta)$ plane along with isocontours for
constant $H$ (highlighted in dark gray is the isocontour corresponding
to the average $H$ and in red is the separatrix delimiting the
area containing asymmetric orbits).}
\label{corot_exp}
\end{figure}

A partial-transfer (5\%) imaging method~\cite{Freilich2010} is employed to create a sequence
of atomic density profiles, as shown in Fig.~\ref{corot_exp}a--d.
The effect of the extractions is primarily to diminish the number of atoms in the
condensate~\cite{Freilich2010,Kuopanportti2011}. Curiously, atomic losses have little
effect on the calculated coupling parameter $c$, which scales only as $\log N$; thus $c$
falls between $0.11$ and $0.10$ over the range $N = 0.3$--$0.8 \times 10^6$ atoms.
For convenience, we take $c=0.1$ in the following analysis.







We examine 52 experimental time series, each consisting of
8 snapshots spanning a total time of 240 to 480 ms. For each
snapshot the center of the vortices and the radius of the
BEC cloud are extracted using a least square fitting algorithm
[see panels a)--d) in Fig.~\ref{corot_exp}].
The vortex positions are then normalized to the BEC radius
(i.e., TF units) and the angular momentum $L_0$ and Hamiltonian
$H$ were computed for each frame [see panels e)--h) in Fig.~\ref{corot_exp}].
For each series, the averaged angular momentum $\overline{L}_0$
and Hamiltonian $\overline{H}$ are computed (see horizontal
dashed lines in the middle panels in Fig.~\ref{corot_exp}).
Using $\overline{L}_0$ we compare the experimental points representing
each orbit in the $(\phi,\delta)$ plane to the isocontour of $H$
corresponding to $\overline{H}$, as shown in the right
column of panels in Fig.~\ref{corot_exp}, and find good
agreement between the two.
%

The panels a)--d) in Fig.~\ref{corot_exp} depict a cohort of
typical time series, together with their respective fits,
that exemplify the different qualitative cases that we observed
in the experiments.
In particular,
we find that the dynamics of the vortices
depends on whether the average
angular momentum is below or above the critical threshold
$L_{{\rm cr}}=2 r_{\rm cr}^2$. This separates
cases where asymmetric orbits are, respectively, non-existent
and possible. The different qualitative cases that we observe,
which are is displayed in Fig.~\ref{corot_exp},
may be grouped as follows:
%
%
\\$\bullet$
For $\overline{L}_0 < L_{{\rm cr}}$ and relatively small
$\overline{H}$, the experiment displays {\em symmetric} orbits.
See rows a) and e) in Fig.~\ref{corot_exp}.
%
\\$\bullet$
For $\overline{L}_0 > L_{{\rm cr}}$ and moderate
$\overline{H}$, the experiment displays
(i) {\em symmetric} orbits where both
vortices (in the co-rotating frame) are approximately on the {\em same}
side of the cloud chasing each other on the same
path [see rows b) and f) in Fig.~\ref{corot_exp}]
or
(ii) {\em asymmetric} orbits [see rows c) and g) in Fig.~\ref{corot_exp}].
The choice between these two orbits is determined by the initial
conditions. Initial conditions inside the area
delimited by the separatrix
(red double-loop curve in the right panels of
Figs.~\ref{corot_exp}.f and Figs.~\ref{corot_exp}.g)
emanating from the saddle point $(\phi,\delta)=(\pi/4,\pi)$
give rise to asymmetric orbits.
%
\\$\bullet$
For $\overline{L}_0 > L_{{\rm cr}}$ and
large $\overline{H}$, the experiment displays orbits in which
one vortex remains close to the center while the other orbits
around it close to the periphery of the cloud.
See rows d) and h) in Fig.~\ref{corot_exp}.

%
%

As 
is clear from these examples and the remaining 48 data sets
that we studied (see supplemental material), asymmetric orbits
are only found when $\overline{L}_0 > L_{{\rm cr}}$ and
when the vortex orbits fall inside the asymmetric minima regions
of the Hamiltonian picture in the $(\phi,\delta)$ plane. 
Asymmetric solutions absent in all of the cases for which
$\overline{L}_0 < L_{{\rm cr}}$.
%
These results are in good agreement with the theoretical prediction
of the pitchfork bifurcation depicted in Fig.~\ref{pde_per}c.

\begin{figure}[t]
\begin{center}
~~~
\includegraphics[width=8.0cm]{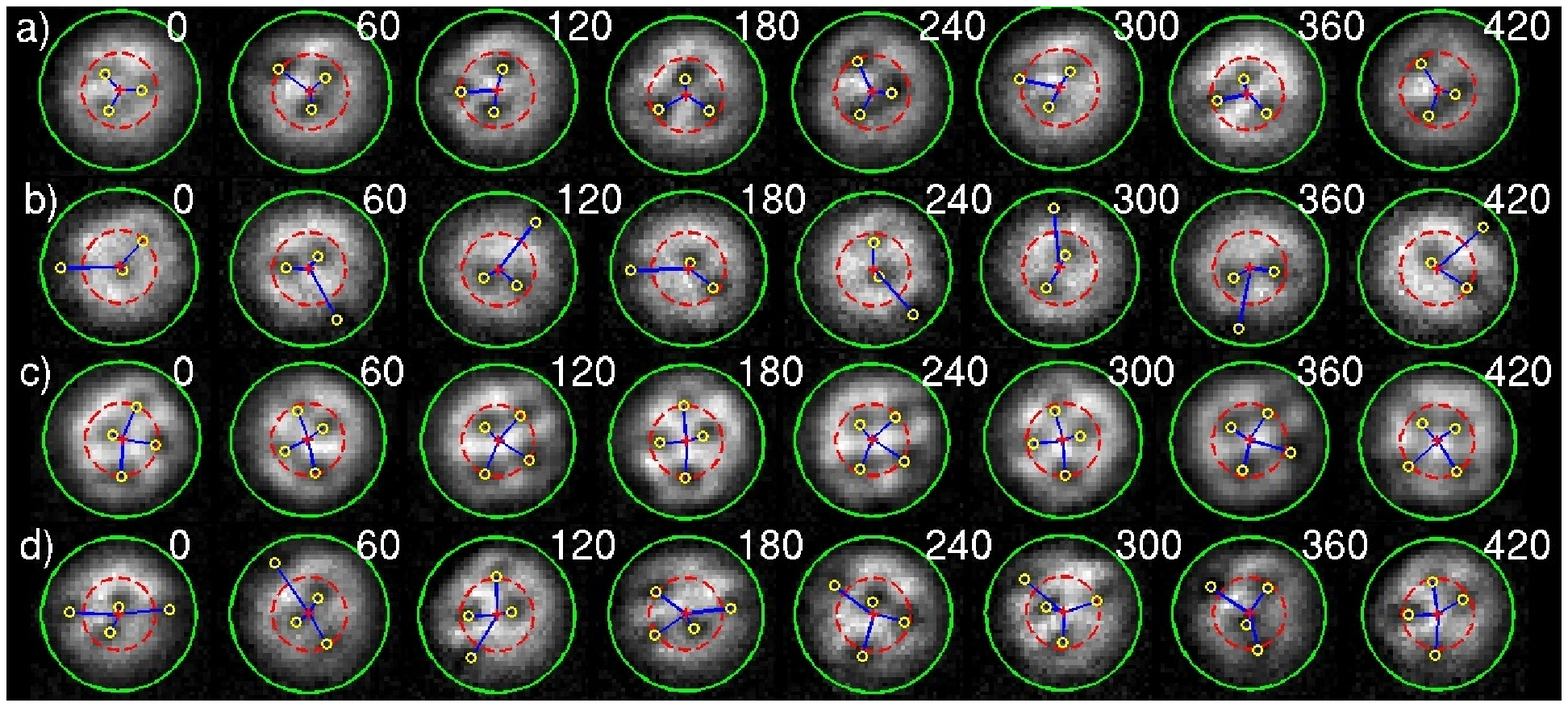}\\[0.5ex]
\includegraphics[height=1.9cm]{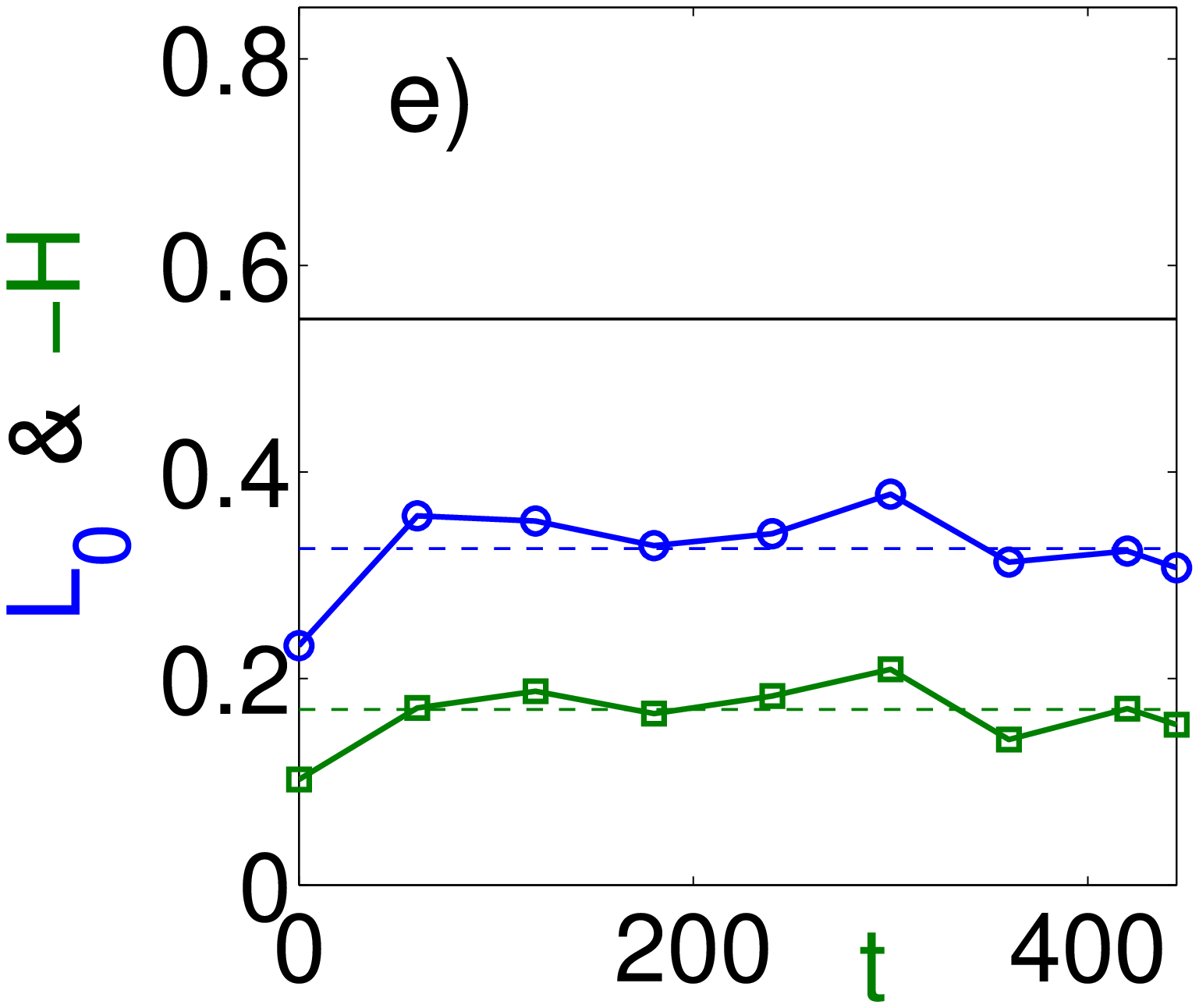}
\includegraphics[height=1.9cm]{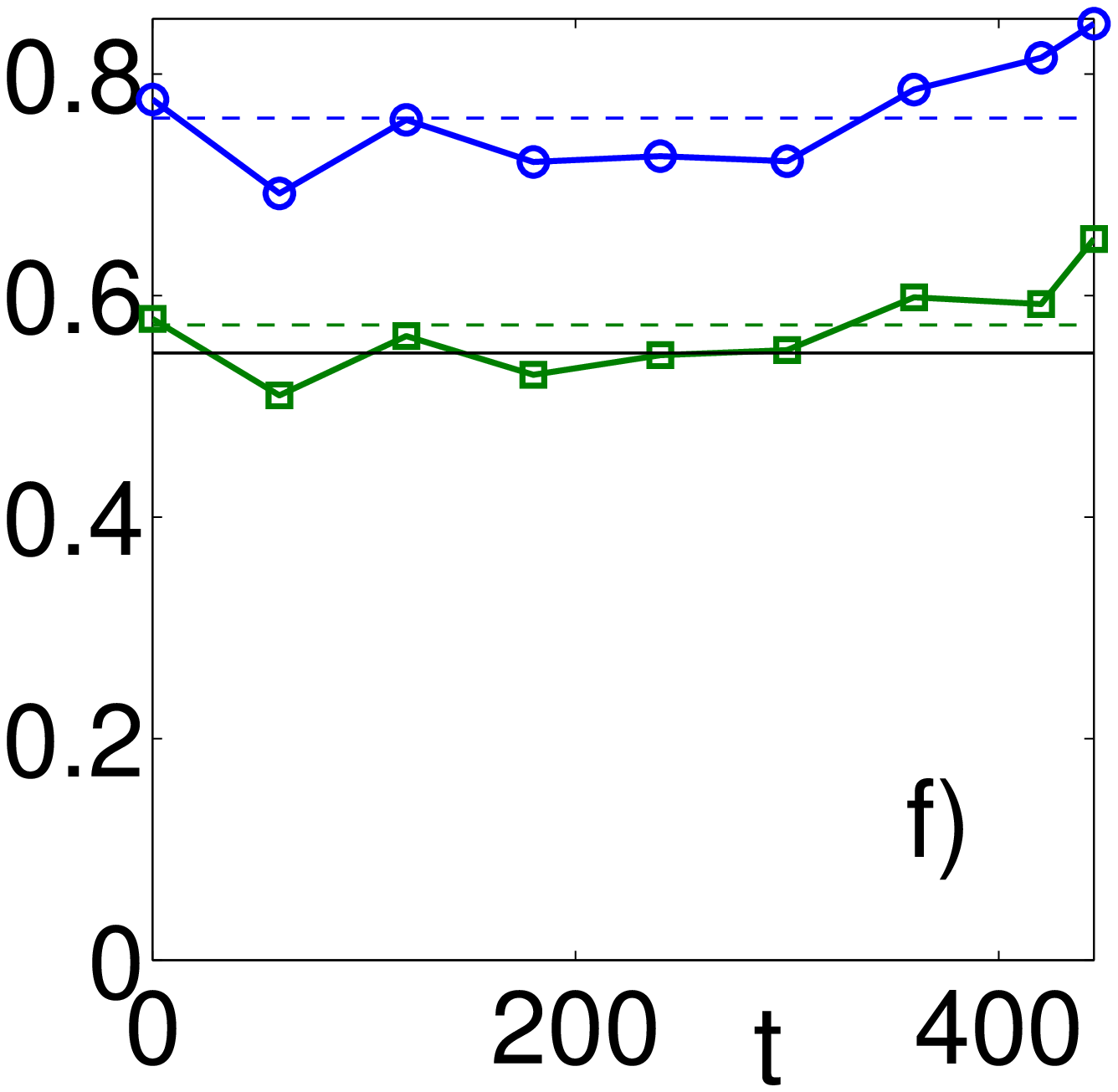}
\includegraphics[height=1.9cm]{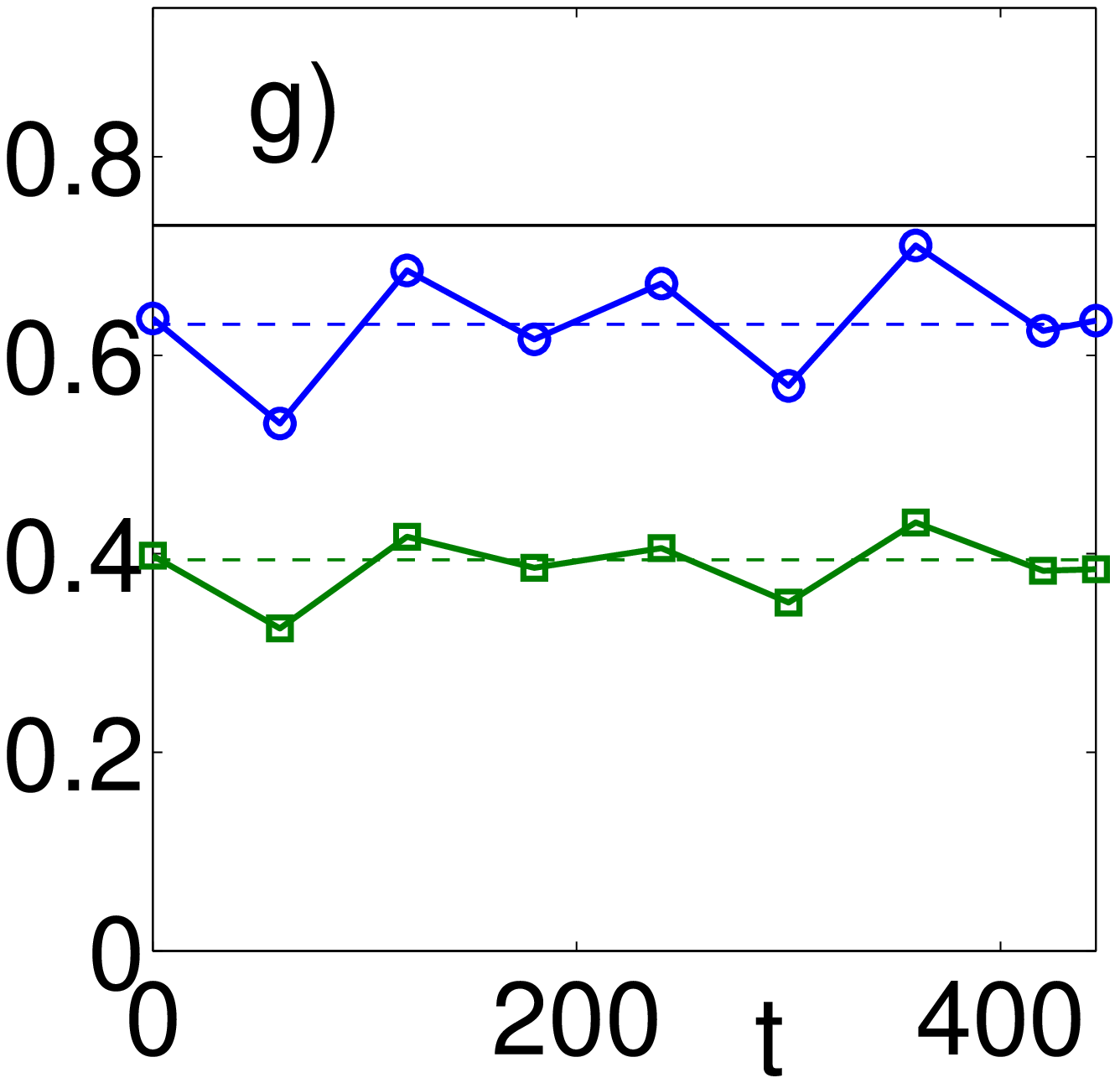}
\includegraphics[height=1.9cm]{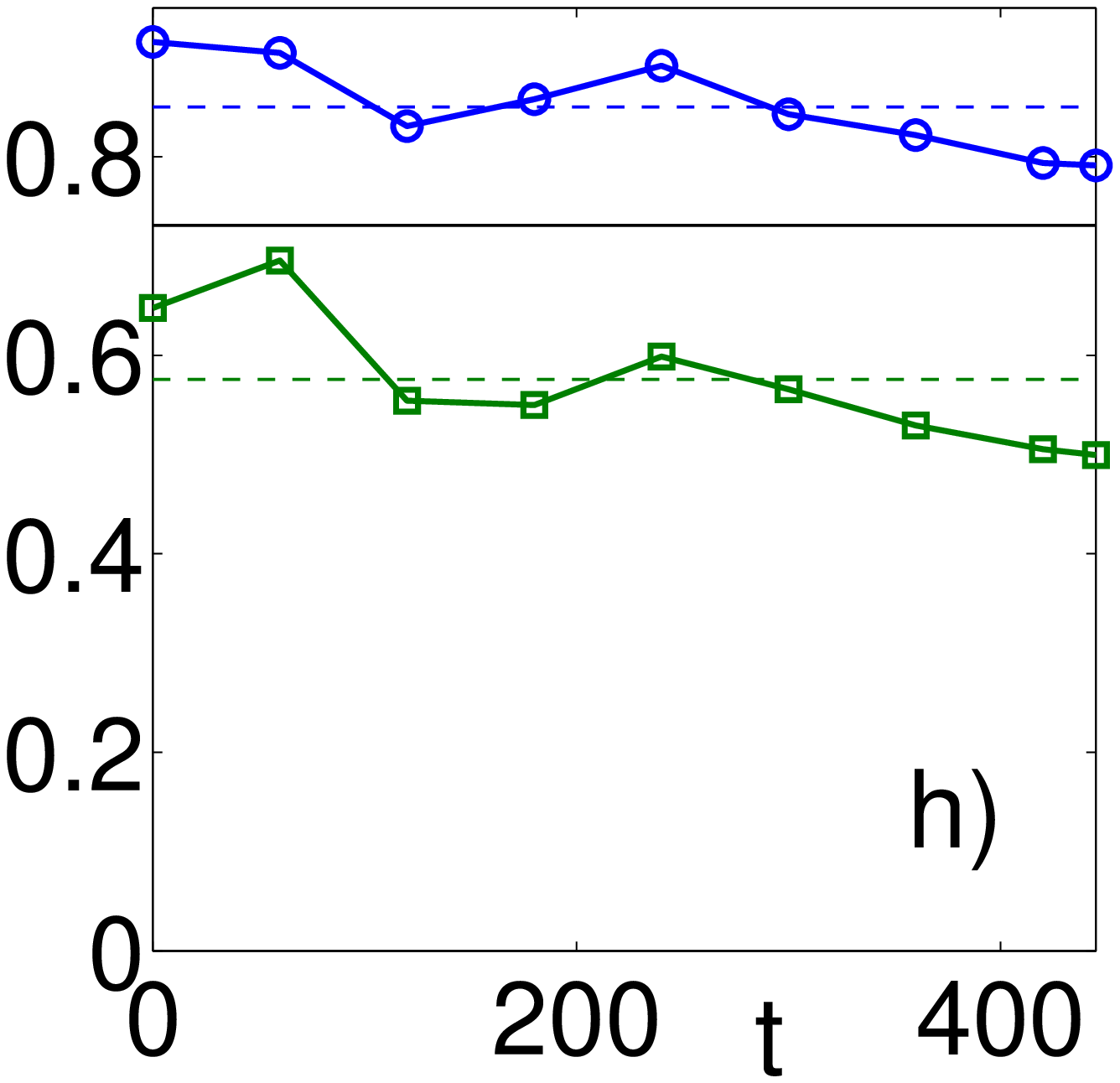}
\end{center}
\par
\vspace{-0.4cm}
\caption{(Color online)
Experimental series for trios [a) and b)] and quartets [c) and d)]
below [a) and c)] and above [b) and d)] the critical threshold.
e)--h) Corresponding time series for $L_0$ and $-H$.
Same notation and units as in Fig.~\ref{corot_exp}.
}
\label{corot_exp_tq}
\end{figure}

To extend our considerations, we briefly present
a comparison between experiment and theory for
$N=3$ and $N=4$
vortices. The main phenomenology is depicted
using two examples for each case in Fig.~\ref{corot_exp_tq}. Panels a) and
b) correspond to the $N=3$ vortex case below and above the
pitchfork bifurcation [see panel e) and f)]. Panels c) and
d) depict the equivalent scenario for $N=4$ vortices.
As
the figure illustrates,
and is observed in all of the cases
that we studied (17 data sets for $N=3$ and 5 data
sets for $N=4$; not shown),
the main phenomenology for $N=3$ and $N=4$ persists in that
all configurations with $\overline{L}_0 > L_{{\rm cr}}$
are not symmetric and symmetric configurations, or
epitrochoidal oscillations about them, are only
present when $\overline{L}_0 < L_{{\rm cr}}$.

{\it Conclusions.}
We have revisited the theme
of co-rotating few vortex clusters in atomic Bose-Einstein
condensates.
By a combination of
theoretical analysis, numerical
computation
and experimental observation, we have illustrated a strong
manifestation of symmetry breaking through a pitchfork bifurcation,
which led to the destabilization of symmetric solidly rotating
configurations
and gave rise to the the emergence of {\em stable} co-rotating
but {\em asymmetric} vortex configurations. We
showed that this analysis is fruitful
not only for the integrable (at the reduced particle level)
two-vortex setting, where a suitable
parametrization of the phase space was provided, but also for
the non-integrable
cases of $N=3$ and $N=4$ vortices
where chaotic orbits exist.

Naturally, it would be interesting to provide a more global
characterization of
the dynamics of the three-body problem, which is perhaps
the most analytically tractable and interesting case due to its potential
for chaos. Another expansion of the present considerations
involves their generalization to higher dimensions. In this case, it would
be interesting to see, upon gradual decrease of the trapping frequency
in the third dimension, whether the symmetry-breaking phenomena
persist 
for line vortices and vortex rings. These aspects
are presently under
study and results will be reported elsewhere.

Acknowledgments: Support from
%
NSF PHY-0855475 (D.S.H.),
%
NSF DMS-0806762 and CMMI-1000337,
and Alexander von Humboldt Foundation (P.G.K.), and
%
NSF DMS-0806762 (R.C.G.),
and discussions with T.K. Langin are kindly acknowledged.


\def\myjump{\hline}
\begin{table}
\begin{tabular}{|c|c|c||c|c|c|}
\myjump
\quad exp\# \quad&\quad~$\overline{L}_0$\quad~&\qquad$\overline{H}$\quad~ &
\quad exp\# \quad&\quad~$\overline{L}_0$\quad~&\qquad$\overline{H}$\quad~ \\[0.1ex]
\myjump
&&&&&\\[-3.2ex]
\myjump
a)  & 0.098 & $\phantom{-}0.0533$ & aa) & 0.219 & $-0.061$ \\\myjump
b)  & 0.117 & $\phantom{-}0.0188$ & bb) & 0.220 & $-0.069$ \\\myjump
c)  & 0.118 & $\phantom{-}0.0175$ & cc) & 0.221 & $-0.033$ \\\myjump
d)  & 0.125 & $\phantom{-}0.0282$ & dd) & 0.224 & $-0.058$ \\\myjump
e)  & 0.131 & $\phantom{-}0.0049$ & ee) & 0.247 & $-0.088$ \\\myjump
f)  & 0.134 & $\phantom{-}0.0062$ & ff) & 0.254 & $-0.085$ \\\myjump
g)  & 0.146 & $          -0.0056$ & gg) & 0.257 & $-0.078$ \\\myjump
h)  & 0.147 & $          -0.0099$ & hh) & 0.257 & $-0.087$ \\\myjump
i)  & 0.153 & $          -0.0125$ & ii) & 0.282 & $-0.111$ \\\myjump
j)  & 0.157 & $          -0.0216$ & jj) & 0.282 & $-0.105$ \\\myjump
k)  & 0.157 & $          -0.0177$ & kk) & 0.300 & $-0.075$ \\\myjump
l)  & 0.164 & $          -0.0162$ & ll) & 0.306 & $-0.137$ \\\myjump
m)  & 0.167 & $          -0.0294$ & mm) & 0.307 & $-0.135$ \\\myjump
n)  & 0.167 & $          -0.0230$ & nn) & 0.333 & $-0.157$ \\\myjump
o)  & 0.169 & $          -0.0215$ & oo) & 0.338 & $-0.166$ \\\myjump
p)  & 0.172 & $          -0.0273$ & pp) & 0.366 & $-0.185$ \\\myjump
q)  & 0.179 & $          -0.0378$ & qq) & 0.373 & $-0.196$ \\\myjump
r)  & 0.184 & $          -0.0404$ & rr) & 0.438 & $-0.241$ \\\myjump
s)  & 0.187 & $          -0.0429$ & ss) & 0.480 & $-0.286$ \\\myjump
t)  & 0.193 & $          -0.0392$ & tt) & 0.519 & $-0.335$ \\\myjump
u)  & 0.194 & $          -0.0517$ & uu) & 0.558 & $-0.372$ \\\myjump
v)  & 0.194 & $          -0.0509$ & vv) & 0.619 & $-0.382$ \\\myjump
w)  & 0.205 & $          -0.0617$ & ww) & 0.655 & $-0.471$ \\\myjump
x)  & 0.208 & $          -0.0634$ & xx) & 0.664 & $-0.489$ \\\myjump
y)  & 0.213 & $          -0.0488$ & yy) & 0.690 & $-0.606$ \\\myjump
z)  & 0.213 & $          -0.0697$ & zz) & 0.751 & $-0.478$ \\
\myjump
\end{tabular}
\caption{Averaged angular momentum $\overline{L}_0$ and
averaged Hamiltonian $\overline{H}$ for all the experimental
series containing two co-rotating vortices. The critical
angular momentum for the experimental setup is
$L_{\rm cr}=0.2731$.}
\label{table}
\end{table}

\begin{figure*}[htb]
\begin{center}
\includegraphics[height=19.0cm]{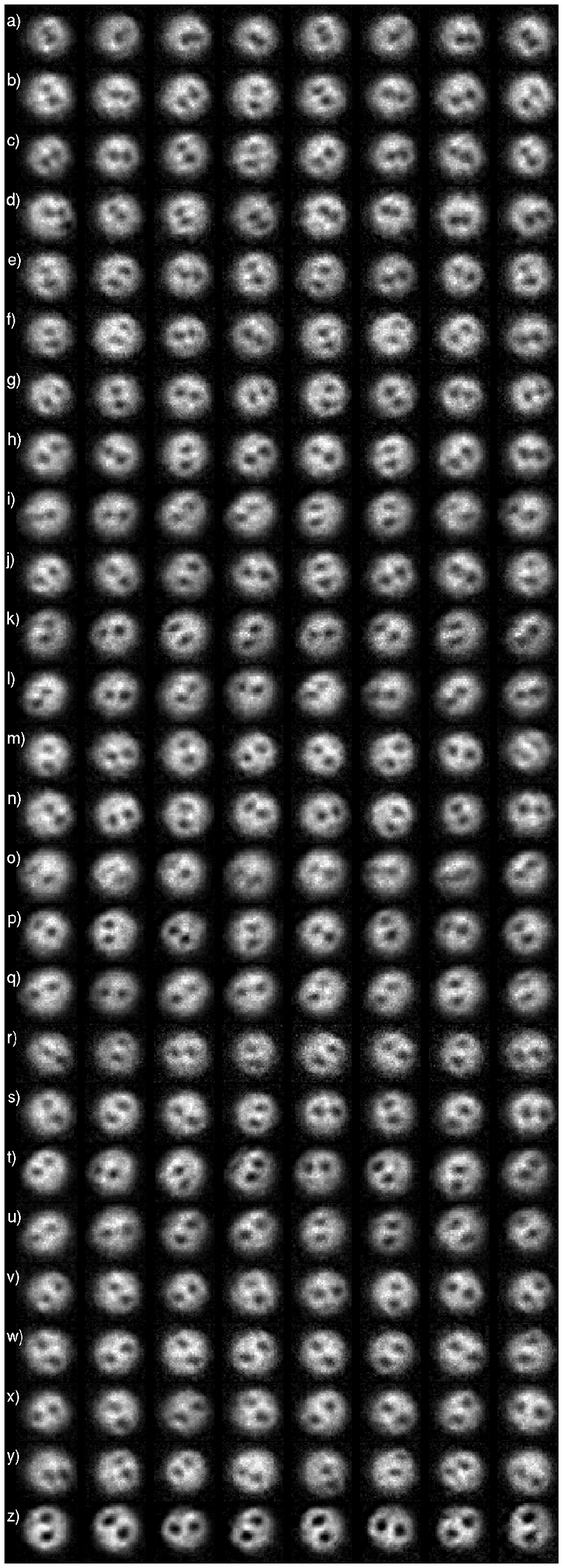}
~~~
\includegraphics[height=19.0cm]{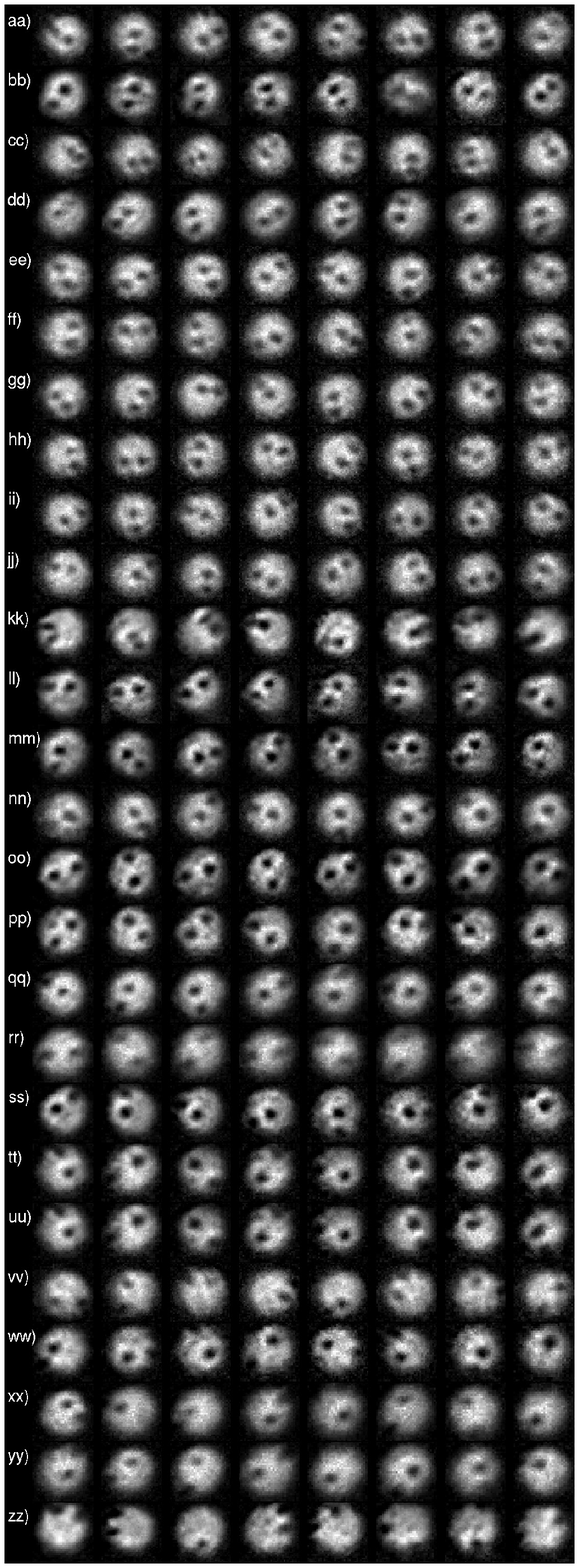}
\end{center}
\vspace{-0.4cm}
\caption{
Experimental time series corresponding to the 52 experiments
containing two co-rotating vortices. The snapshots were taken
at regular times for a total time spanning 240 ms or 480 ms
(for individual times for each snapshot refer to the corresponding
middle column in Fig.~\ref{fig:supp_L0}).
The experiments are ordered according to average angular
momentum as listed in Table~\ref{table}.
For ease of presentation,
all the images have been centered and rescaled so that they
occupy approximately the same visual area.
}
\label{fig:supp_snaps}
\end{figure*}

\begin{figure*}[t]
\begin{center}
\includegraphics[width=17.9cm]{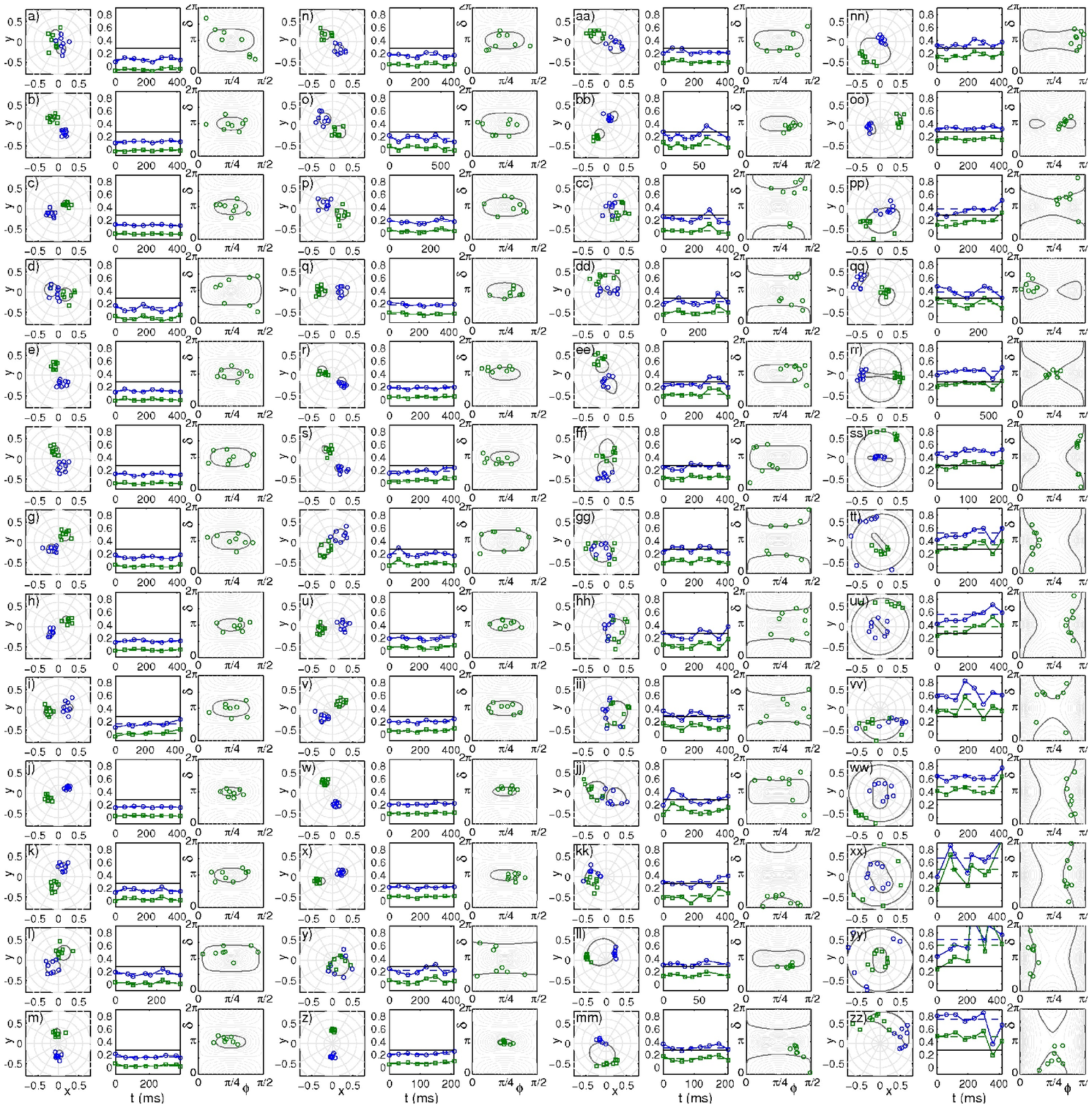}
\end{center}
\vspace{-0.4cm}
\caption{(Color online)
Analysis for experimental series depicted in Fig.~\ref{fig:supp_snaps}.
For each series the respective columns represent:
(i) The left column depicts the vortex orbits
(green squares and blue circles),
and their corresponding orbit from the
reduced ODE model (gray solid line), in TF units in the
co-rotating frame.
(ii) The middle column depicts the time series for
the angular momentum $L_0$ (blue circles) and
negative Hamiltonian $-H$
(green squares) and their respective averages
(horizontal dashed lines) as well as
the critical value for $L_0$ (solid horizontal line).
(iii) The right column depicts the orbits in the
$(\phi,\delta)$ plane along with the isocontours for
constant $H$ (highlighted is the isocontour corresponding
to the average $H$).
}
\label{fig:supp_L0}
\end{figure*}

\bigskip
\centerline{\bf SUPPLEMENTAL MATERIAL:}

Here we  present the results that we obtained for the 52 experimental
series containing two co-rotating (same charge) vortices.
The experiments are ordered by
the average angular momentum $\overline{L}_0$.
which is
computed from the fitted centers and cloud diameter.
The average is then computed by averaging over the eight
experimental snapshots for each series.
Table~\ref{table} lists the average angular momentum and average
Hamiltonian for all the experimental series containing
two co-rotating vortices.
The corresponding snapshots for all the series are shown in
Fig.~\ref{fig:supp_snaps} using the same ordering as given
in Table~\ref{table}.
Finally, Fig.~\ref{fig:supp_L0} depicts the analysis for
each individual series by presenting the orbits in the co-rotating
frame (respective left columns), the times series for the
angular momentum and Hamiltonian (respective middle columns), and
the orbits in the reduced $(\phi,\delta)$ plane (respective
right columns).

As observed in Fig.~\ref{fig:supp_L0}, the experimental
orbits in the co-rotating frame (respective left columns)
match well the theoretical orbits from our model
(see gray orbit) for all of the experimental runs.
%
Significantly also, all experimental orbits
in the reduced $(\phi,\delta)$ plane (respective right columns)
match well the theoretical orbits corresponding to the
isocontour of constant $H$ equal to the average $H$ from the
experiment ---with the exception of the experiment\#rr which we discuss
below.
In most of the experimental series both the angular momentum and the
Hamiltonian are approximately conserved, buttressing the validity of our
dynamical reduction.
%

To further bolster the case for the validity of
our reduced dynamics we now concentrate
on the main result of our manuscript.
The reduced model predicts that,
as the angular momentum
is increased, the co-rotating symmetric configuration loses
stability in a pitchfork, symmetry-breaking
bifurcation. At the bifurcation point
a pair of stable, asymmetric co-rotating configurations
emerge. Here, the bifurcation point is predicted to occur at
$L_{\rm cr}=2 r_{\rm cr}^2$ where the square of the critical
radius is $r_{\rm cr}^2 = \sqrt{c}/(\sqrt{c}+2)$
in units of the Thomas-Fermi radius.
%
In our experiment, the value of $c$, the ratio
between the rotation induced by vortex-vortex interactions
and the rotation induced by the precession of vortices
due to the inhomogeneity of the cloud's background
produced by the external magnetic trap, is
$c \approx 0.1$. This gives a value for the
critical angular momentum in our experiments
of $L_{\rm cr}=0.2731$.
Therefore, corroboration of our analytical prediction
about the pitchfork bifurcation
would imply that (i) for all the experimental series
with an average angular momentum below critical
($\overline{L}_0 < L_{\rm cr}$) there should be
no co-rotating asymmetric orbits, and (ii) if an
asymmetric co-rotating orbit is observed its
corresponding angular momentum should be above
the critical threshold.
The critical threshold for the angular momentum
lies between experiments\#hh and ii.

It should be noted at this stage that an angular momentum
above critical does not imply an asymmetric co-rotating
orbit. For example orbits with high angular momentum
(see experiments\#ss--zz) might correspond to (i) orbits where
vortices seem decoupled and one vortex remains close to
the trap center while the other orbits around close to the
periphery of the cloud (see experiments\#ss, tt, uu, ww,
xx, yy), or (ii) the two vortices belong to the
same path that has them on the same side of the cloud
(see experiments\#vv and zz).
The type of resulting orbit is heavily dependent on the initial
conditions.
In fact, in the $(\phi,\delta)$ plane, the area of the
set of initial conditions that leads
to co-rotating asymmetric orbits (or slight perturbations
thereof) is rather small and therefore asymmetric
co-rotating instances are less common.
Nonetheless, we were able to identify two orbits as
being asymmetric (i.e., epitrochoidal perturbations
of the asymmetric orbit) in experiments\#oo and qq. In fact
experiment\#oo is featured in
panels c) and g) in Fig.~3 of the manuscript.
It is reassuring that the angular momenta corresponding
to the experimental asymmetric cases is clearly
above critical supporting our
dynamical reduction and the pitchfork bifurcation picture
that it unveiled.

It is important to note that one of the 52 experimental
series did not match our theoretical
expectations. This case is experiment\#rr.
As it is clear from the right panel of Fig.~\ref{fig:supp_L0}.rr,
the experimental orbit lies close to the
point $(\phi,\delta)=(\pi/4,\pi)$ indicating an
almost perfect {\em symmetric} orbit as observed in
the left panel of Fig.~\ref{fig:supp_L0}.rr. However,
this case has an average angular momentum of
$\overline{L}_0=0.438$, well above the
critical threshold $L_{\rm cr}=0.2731$.
The reduced dynamics description predicts that this
orbit should be unstable, lying as it does close to
an unstable saddle.
It is not clear how this symmetric orbit can exist in
this supposedly unstable
region.
%
It is possible, albeit unlikely,
that
this is an unstable orbit
that has not had enough time for the instability to
develop.
A second, more plausible, explanation can be given if
one carefully examines the experimental snapshots
corresponding to this experiment depicted in
Fig.~\ref{fig:supp_snaps}.rr. It is evident that in this
experimental series, the traces of the vortices are much
more diffuse and blurry than in the other series.
%
The vortex lines corresponding
to the vortices in this series may not be straight or aligned with
the trap axis
but rather bent \cite{VR:USshapedVLs1,VR:USshapedVLs2}.
This explains on one hand the blurriness of the vortex
cores in the experimental snapshots and the unusual
(with respect to our analytical dynamical reduction)
dynamics that they display.
This would further serve to illustrate the consequence of
the true three-dimensional
nature of the condensate that, in some instances, cannot
be captured by our reduced {\em two-dimensional} description.
%



\begin{thebibliography}{99}

\bibitem{bec}
C.J. Pethick and H. Smith,
Bose-Einstein Condensation in Dilute Gases, Cambridge University Press (2002);
L. Pitaevskii and S. Stringari, Bose-Einstein Condensation, Oxford University Press, New York (2003).

\bibitem{castin}
Y. Castin and R. Dum,
Eur. Phys. J. D, {\bf 7}, (1999) 399.

\bibitem{Fetter2001}
A.L. Fetter and A.A. Svidzinsky,
J. Phys.: Condens. Matter, {\bf 13} (2001) R135.

\bibitem{Kevrekidis2008}
P.G. Kevrekidis, D.J. Frantzeskakis and R. Carretero-Gonz{\'a}lez,
Emergent Nonlinear Phenomena in Bose-Einstein Condensates,
Springer-Verlag, Berlin (2008).

\bibitem{Fetter2009}
A.L. Fetter,
Rev. Mod. Phys., {\bf 81} (2009) 647.

\bibitem{Newton2009}
P.K. Newton and G. Chamoun,
SIAM Review {\bf 51}, 501 (2009).

\bibitem{Crasovan2002}
L.-C. Crasovan,
{\it et al.},
Phys. Rev. E, {\bf 66} (2002) 036612.

\bibitem{Crasovan2003}
L.-C. Crasovan,
{\it et al.},
Phys. Rev. A, {\bf 68} (2003) 063609.

\bibitem{Zhou2004}
Q. Zhou and H. Zhai,
Phys. Rev. A, {\bf 70} (2004) 043619.

\bibitem{Mottonen2005}
M. M{\"o}tt{\"o}nen,
{\it et al.},
Phys. Rev. A, {\bf 71} (2005) 033626.

\bibitem{Pietila2006}
V. Pietil{\"a},
{\it et al.},
Phys. Rev. A, {\bf 74} (2006) 023603.

\bibitem{Li2008}
W. Li,
{\it et al.},
Phys. Rev. A, {\bf 77} (2008) 053610.

\bibitem{Middelkamp2010}
S. Middelkamp,
{\it et al.},
Phys. Rev. A, {\bf 82} (2010) 013646.

\bibitem{Kuopanportti2011}
P. Kuopanportti,
{\it et al.},
Phys. Rev. A, {\bf 83} (2011) 011603.

\bibitem{Torres2011}
P.J. Torres,
{\it et al.},
Phys. Lett. A {\bf 375}, 3044 (2011).

\bibitem{Neely2010}
T.W. Neely,
{\it et al.},
Phys. Rev. Lett. {\bf 104 }(2010) 160401.

\bibitem{Freilich2010}
D.V. Freilich,
{\it et al.},
Science, {\bf 329} (2010) 1182.

\bibitem{Seman2010}
J.A. Seman,
{\it et al.},
Phys. Rev. A, {\bf 82} (2010) 033616.

\bibitem{Middelkamp2011}
S. Middelkamp,
{\it et al.},
Phys. Rev. A {\bf 84}, 011605(R) (2011).

\bibitem{Afta03_04_Danaila05}
A. Aftalion and I. Danaila.
Phys. Rev. A, {\bf 68} (2003) 023603;
%
{\it ibid.}
Phys. Rev. A, {\bf 69} (2004) 033608;
%
I. Danaila.
Phys. Rev. A, {\bf 72} (2005) 013605.

\bibitem{komineas_rev}
S. Komineas,
Eur. Phys. J.-Spc. Top. {\bf 147}, 133 (2007).

\bibitem{pelin2011}
D.E. Pelinovsky and P.G. Kevrekidis,
Nonlinearity {\bf 24}, 1271 (2011).

\bibitem{screening}
S. McEndoo and Th. Bush,
Phys. Rev. A, {\bf 79} (2009) 053616.

\bibitem{Hodby2001}
E. Hodby,
{\it et al.},
Phys. Rev. Lett. {\bf 88} (2001) 010405.

\bibitem{Madison2000}
K.W. Madison,
{\it et al.},
Phys. Rev. Lett. {\bf 84} (2000) 806. 


\bibitem{VR:USshapedVLs1} 
A. Aftalion and I. Danaila,
Phys.\ Rev.\ A {\bf 68} (2003) 023603.

\bibitem{VR:USshapedVLs2} 
S. Komineas, N. R. Cooper and N. Papanicolaou,
Phys.\ Rev.\ A {\bf 72} (2005) 053624.




\end{thebibliography}
\end{document}